\documentclass[review]{elsarticle}
\usepackage{subfigure,dcolumn}
\usepackage[T2A,T1]{fontenc}
\usepackage{multirow} 
\usepackage{amsmath}
\usepackage{booktabs}
\usepackage[russian,english]{babel}

\usepackage{listings}
\usepackage{lineno,hyperref}
\modulolinenumbers[1]

\journal{Nucl. Instrum. Methods. Phys. Res. A}
\usepackage{bm}
\usepackage{subfigure,dcolumn}

\begin{document}

\begin{frontmatter}

\title{The design of high-brightness ERL-FEL injector based on VHF electron gun\tnoteref{mytitlenote}}

\author[mymainaddress,mysecondaryaddress]{Xiuji Chen}
\author[myfirdaddress]{Zipeng Liu\corref{mycorrespondingauthor1}}\ead{liuzp@zjlab.ac.cn}
\author[mymainaddress]{Xuan Huang}
\author[mythirdaddress]{Si Chen}
\author[mythirdaddress]{Duan Gu}
\author[myfirdaddress]{Houjun Qian}
\author[mythirdaddress]{Dong Wang}
\author[mythirdaddress]{Haixiao Deng\corref{mycorrespondingauthor2}}\ead{denghx@sari.ac.cn}




\address[mymainaddress]{Shanghai Institute of Applied Physics, Chinese Academy of Sciences, Shanghai 201800, China}
\address[mysecondaryaddress]{ShanghaiTech University, Shanghai 201210, China}
\address[myfirdaddress]{Zhangjiang Laboratory, Shanghai 201200, China}
\address[mythirdaddress]{Shanghai Advanced Research Institute, Chinese Academy of Sciences, Shanghai 201210, China}

\begin{abstract}
In the past decade, the fourth-generation light source based on the combination of Energy Recovery Linac (ERL) and Free-Electron Laser (FEL) using superconducting linear accelerators has garnered significant attention. It holds immense potential, particularly in generating high-power Extreme Ultraviolet (EUV) light sources. This article primarily focuses on the physical design of an injector for ERL-FEL, based on a Very High Frequency (VHF) electron gun with a charge of 100 pC. The beam energy is accelerated to 10 MeV using 3-cell superconducting cavity. The optimization of beam parameters is conducted through employment of BMad and ASTRA simulations, incorporating the concept of Merger optimization. The beam emittance is less than 0.6 mm mrad, and the peak current at the injector exit exceeds 18 A. We present a new method to evaluate the Longitudinal Space Charge (LSC) effects in merger sections, which can be readily applied in design work. Furthermore, we introduce a novel type of merger. The performance of this new merger is comparable to the previously known optimum, the zigzag merger, offering a potential alternative solution for injectors in ERLs.
\end{abstract}


\end{frontmatter}


\section{Introduction}
 In the past decades, the free electron laser (FEL) have emerged as indispensable tools in numerous high-tech domains such as materials science, surface science, atomic and molecular physics, and many others \citep{Ilchen2018,Kortright1999}. To accommodate the operation of additional experimental beamlines and increasing average laser power, FEL facility necessitate higher repetition rates. Considering the high average power of beam recovery, the combination of an Energy Recovery Linac (ERL) with a FEL is regarded as a strong contender for the next-generation light source\cite{Williams2002}. Some theoretical design and experimental research related to ERL-FEL have been carried out by research institutes and universities. Additionally, the facilities based on high-energy, high-average current ERLs have also been proposed and used to generate high-power, high-brightness FEL light sources.\cite{Gulliford2013,Nakamura2023,Jones2009,Vinokurov2018}. In ERL systems, the injector plays a crucial role as a highly important component that determines the performance of the device. The beam generated by the ERL will ultimately be used to generate high-power laser through the FEL. Therefore, the  emittance and bunch length will have an impact on the efficiency of light generation. To meet these requirements, it is necessary to reduce the growth of beam emittance caused by space charge effects. The peak current is a critical requirement for generating high-power FEL. However, the electron beam produced in the injector has an extended bunch length, necessitating compression to the kA level in subsequent compression stages. The efficiency of beam compression relies on the structure and shape of the longitudinal phase space, with higher-order terms playing a significant role in determining downstream compression efficiency \cite{Floettmann2014,Krasilnikov2007}. Therefore, this article also focuses on non-linear longitudinal phase space optimizing to improve the efficiency of the compressor.

Research efforts have been carried out on the theoretical and experimental aspects of injectors for ERL systems. These efforts primarily include studies on injectors based on DC electron guns and 2-cell injector cryomodule\cite{Miyajima2015,Gulliford2013,Kewisch2011}. DC electron guns are characterized by lower acceleration gradients, and the space charge effects have a significant impact on the beam quality of the electron bunch. A Superconducting Radio Frequency (SRF) electron gun has been 
designed, which operates at a frequency of 704 MHz and achieves an acceleration of up to 2 MeV in continuous wave operation mode\cite{Vennekate}. A 1.4$\cdot \lambda/2$ cell SRF photoinjector cavity with $K_2CsSb$ cathode was commissioned at bERLinPro\cite{Neumann2017}, and the injector designed average current of 6 mA at up to 3.5 MeV kinetic energy. Considering that the usage of a microwave electron gun results in superior beam quality compared to a DC electron gun, and the complexity of low-temperature superconducting systems. Additionally, we take into account that high-repetition-rate FEL facilities based on Very-High-Frequency electron guns are currently under construction and in operation such as SHINE\citep{Zhu2017} and LCLS-II\cite{Stohr2011,Zhang2023}.  This article adopts a VHF electron gun and 3-cell injector cryomodule to generate a high-quality beam for the ERL system. The electron gun operates at a frequency of 216.667 MHz, and the accelerating gap is 40 mm. Additionally, considering the characteristics of high current, semiconductor photocathodes with high quantum efficiency should be chosen,  Considering the high-average-current  ERL system and the current quantum efficiency of traditional semiconductor photocathodes, a high-power driving laser is needed. Taking into account the stability of the driving laser, there is a preference for using "green" cathodes as the photocathodes. 

For machines equipped with a circular beamline, the integration of a merger section is crucial for enabling the seamless transition of the beam from the injector to the circulator. In ERL systems, the beam energy upon exiting the injector typically remains below 20 MeV to optimize energy efficiency. As a result, the bunches within the merger operate within the space charge regime, where they are notably affected by space charge forces. This influence can result in a degradation of beam quality.

Unlike the straight segments found in the injector or matching line, the merger operates as a deflection line characterized by non-zero dispersion. This design feature introduces additional effects on the bunches. Specifically, the variability in energy spread, driven by longitudinal space charge forces (LSCF, or simply LSC in this article), can significantly impact beam dynamics. When designing merger sections for low-brightness machines, spatial constraints within the accelerator facility often take precedence. Key design parameters—including the length of the merger, the specific deflection angles, and the precise positioning of dipole magnets—must be meticulously optimized. In some cases, a large total deflection angle is necessary to effectively guide the bunches into the circulator. However, this requirement can lead to considerable degradation of beam quality within the merger section\cite{Holder2009, Vinokurov2008}. 
In most cases, a merger section that is both short in length and simple in structure is preferred. This preference arises not only from the ease of engineering implementation but also due to the presence of other collective effects, such as transverse space charge forces (TSCF), which can lead to additional transverse emittance growth across the entire space charge regime. In some instances, this growth may exceed that caused by LSC \cite{Cee2002, Mizuno2015}.
A more sophisticated approach involves the implementation of a zigzag-type merger \cite{Kayran2005, Litvinenko2006}, as illustrated in Fig. (\ref{fmer1}). This design features a straightforward structure and effectively eliminates first-order dispersion while mitigating LSC effects. Comparisons in Ref \cite{Litvinenko2006} demonstrate the strong capability of zigzag merge to suppress emittance growth. However, the compact spatial structure of the zigzag merger presents challenges for practical implementation in existing machines. In this work, given the high beam intensity and low energy at the injector exit, it is essential to effectively suppress emittance growth caused by LSC effects, as lower transverse emittance significantly enhances the power of FEL.

The paper is organized as follows. Section \ref{S1A} focuses on the design of the injector, providing a detailed overview of the underlying principles and optimized methods. In Section \ref{S2B}, we explore the design of the merger, introducing a novel approach. The effectiveness of this new method is qualitatively demonstrated and subsequently applied to the design of a Triple Bend Achromat (TBA) merger. Section \ref{S3} addresses the matching section between the injection and merger sections. Finally, start-to-end (S2E) simulations are conducted to validate the performance and integration of the entire system.

\section{Injector with Merger section}\label{sec:artwork}

\begin{figure}[htpb!]
    \centering
    \subfigure{\includegraphics*[width=1\linewidth]{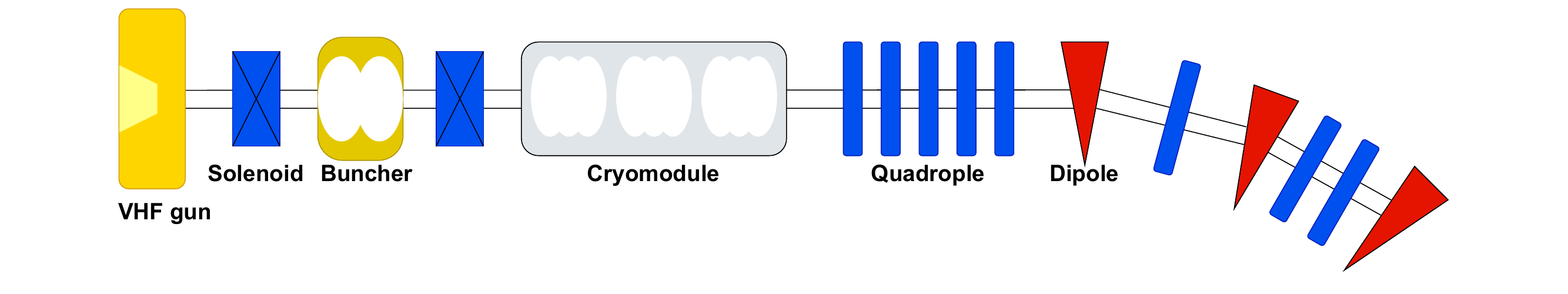}}
    \caption{The main layout of ERL injector.}
    \label{fig:ERL_layout}
\end{figure}

A typical injector section for an ERL comprises a conventional linac injector (referred to as the "injection section" in this study), a matching section, and a merger section. The layout of the ERL injector is illustrated in Fig.~(\ref{fig:ERL_layout}). The primary purpose of the matching section is to control the bunch envelope from the exit of the injection section to the entrance of the merger. Consequently, this section focuses exclusively on the injection and merger sections.
High-brightness bunches are essential for the ERL-FEL injector. Bunches with lower transverse emittance can significantly enhance the power of FEL \cite{chao2023handbook}. Additionally, high peak current is required for effective FEL processes. An initially shorter bunch length aids in preserving emittance during the bunch compression process \cite{DiMitri2016}. Therefore, our design work prioritizes achieving shorter bunches with lower transverse emittance.
\subsection{Injection Design} \label{S1A}
\begin{figure}[htpb!]
    \centering
    \subfigure{\includegraphics*[width=0.9\linewidth]{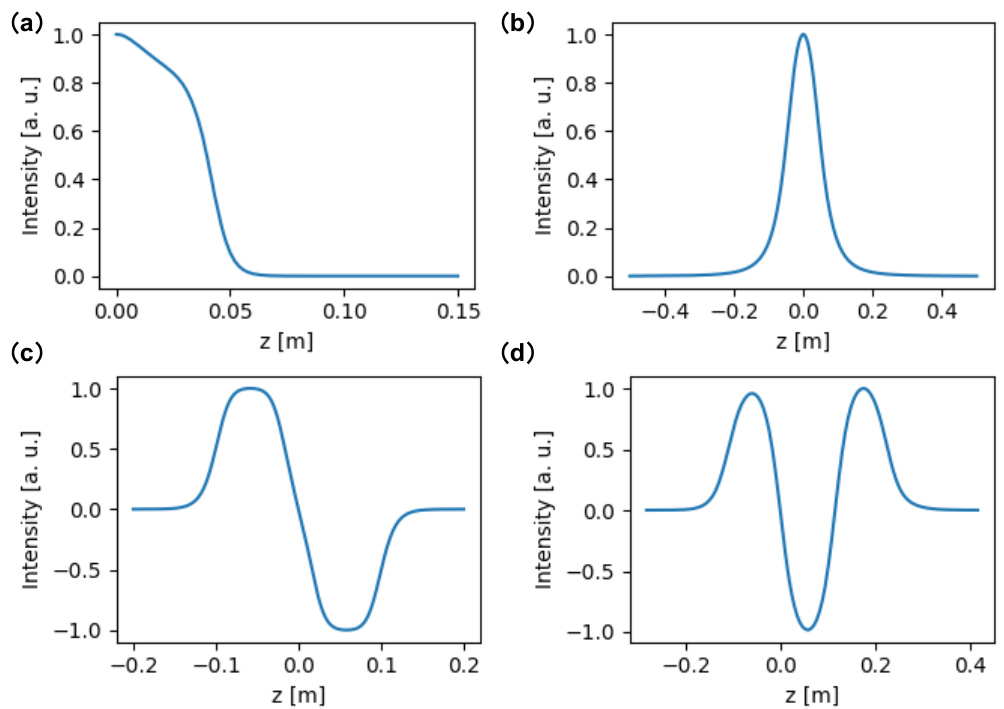}}
    \caption{The longitudinal electric (a: electron gun, c: buncher, d: 3 cell cavity) and magnetic (b: solenoid) field of electron gun, solenoid, buncher and cavity.}
    \label{fig:RF_layout}
\end{figure}

The injection section comprises a VHF electron gun, two solenoids, a 1.3 GHz buncher, and a cryomodule containing three 3-cell cavities. The field strengths of the VHF electron gun, buncher, solenoids, and 3-cell cavities are illustrated in Fig.~(\ref{fig:RF_layout}). The 216.667 MHz VHF electron gun has been designed and successfully commissioned at the Shanghai Advanced Research Institute (SARI)\cite{Liu2023}. This electron gun features an accelerating gap of 40 mm, with an estimated total RF input power of approximately 70 kW required to accelerate the beam to 720 keV.
The normalized emittance is a critical parameter for FEL lasing, and to mitigate the growth in emittance at the exit of the gun, a solenoid is employed. The superconducting accelerator module, which consists of the 3-cell cavities, is utilized to accelerate the electron beam to about 10 MeV over a short distance. Subsequently, the beam is injected into the ERL ring through the merger section.

In contrast to the optimization methods typically employed for conventional linear accelerators, optimizing an ERL injector requires a dual focus on the  requirements of mergers. To mitigate the impact of projected emittance resulting from space charge effects, a flat-top laser duration of 20-40 ps was utilized to optimize the layout and parameters of the injector \cite{Krasilnikov2007}. Additionally, a buncher is essential for compressing the bunch at the exit of electron gun.
The injection section comprises several RF cavities, including the VHF gun, a 1.3 GHz buncher, and three 1.3 GHz 3-cell accelerating units (designated as CAV01-03). Two solenoids (labeled SOL01 and SOL02) are also incorporated. The optimization of the layout and parameters of the injection section was conducted using a combination of ASTRA \cite{KFloettmann} and the Non-dominated Sorting Genetic Algorithms (NSGA-II) \cite{Deb2002}.
Key parameters considered in this optimization include the duration and transverse diameter of the driving laser (with rising and falling edges of 2 ps), the positions of the RF cavities (GUN, BUN, CAV01-03), and solenoids (SOL01, SOL02), along with the phase and amplitude of the RF cavities and the strength of the solenoids. The thermal emittance of the cathode is set to $1\ mm\cdot mrad/mm$, based on research conducted on $Cs_2Te$ photocathodes, which exhibited a measured thermal emittance of 1 $mm\cdot mrad/mm$.

\begin{figure}[htpb!]
    \centering
    \subfigure{\includegraphics*[width=0.9\linewidth]{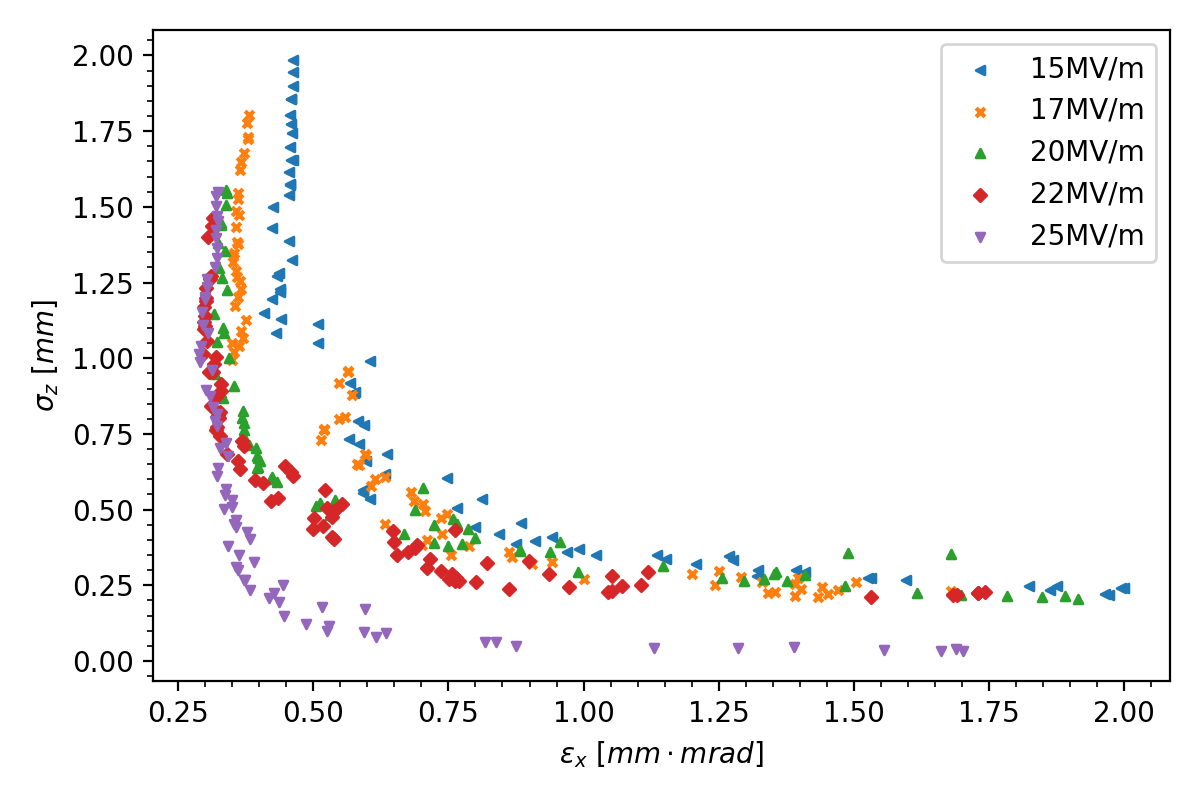}}
    \caption{The figure represents the correlation between bunch charge and normalized emittance with different electron gun gradient of 15, 17, 20, 22 and 25 MV/m, and the bunch charge is 100 pC. The dashed red line represents the normalized emittance with the bunch length of 0.8 mm.}
    \label{fig:difEGun}
\end{figure}

\begin{figure}[htpb!]
    \centering
    \subfigure{\includegraphics*[width=0.9\linewidth]{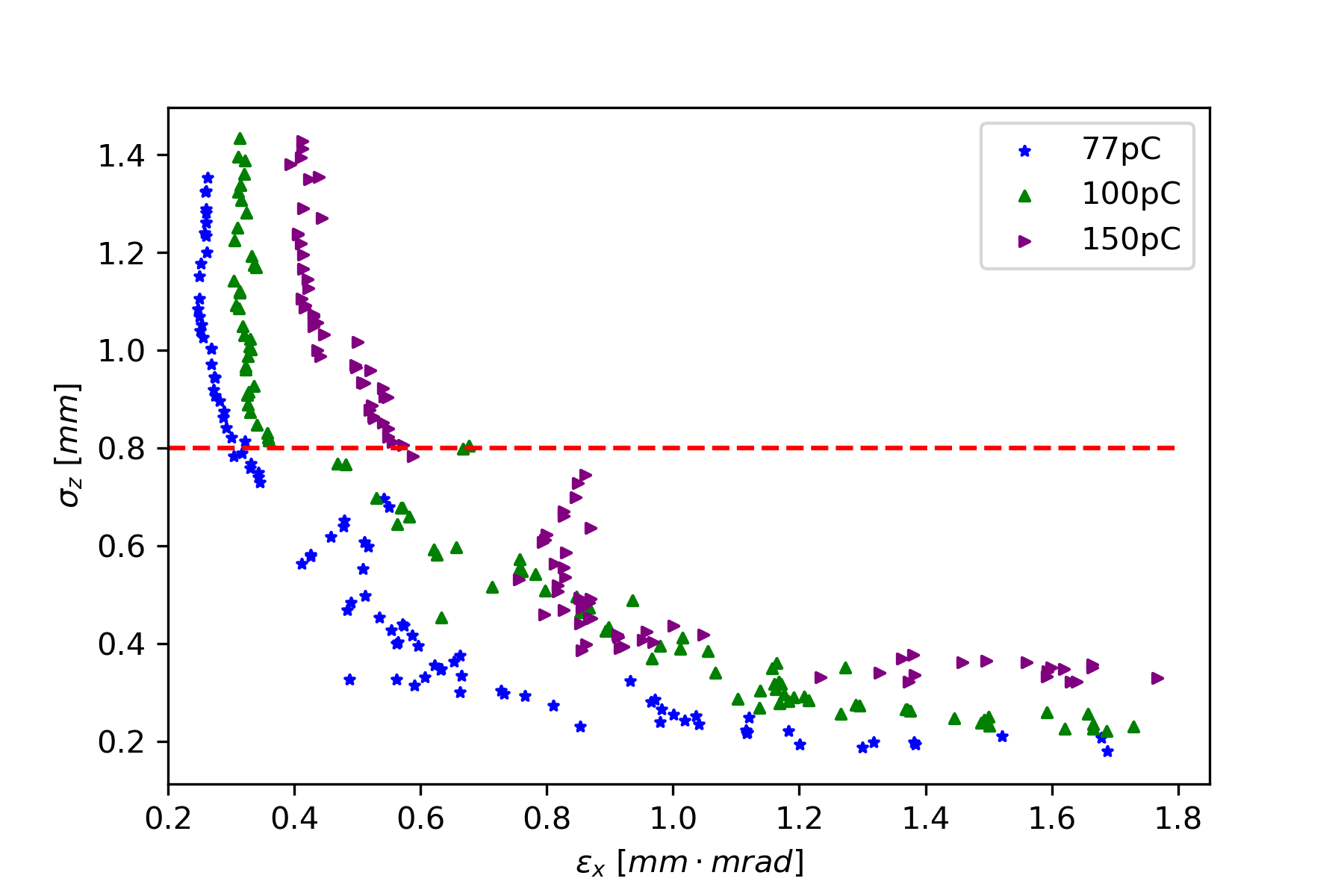}}
    \caption{The figure represents the correlation between bunch length and normalized emittance with different bunch charge of 77, 100 and 150 pC.}
    \label{fig:difCharge}
\end{figure}

We accounted for the effects of beam normalized emittance, bunch length, peak current and longitudinal phase space. As illustrated in Fig.~(\ref{fig:difEGun}), we simulated the relationship between beam emittance and bunch length at the exit of the module under acceleration gradients of 15, 17, 20, 22, and 25 MV/m. The dashed red line represents the beam emittance for a bunch length of 0.8 mm across different acceleration gradients. Considering the stability of the electron gun and the efficiency of FEL radiation, we ultimately selected 20 MV/m as the reference acceleration gradient. As previously mentioned, the VHF electron gun has operated in CW mode, with high current mode currently under commissioning. Based on the 20 MV/m electron gun acceleration gradient, we conducted further simulations to explore the relationship between beam emittance and bunch length for various bunch charges, as shown in Fig.~(\ref{fig:difCharge}). We identified a representative charge value of 100 pC as the optimal operating point.
As mentioned previously, seventeen parameters of the beamline have been selected as the optimized variables, as detailed in Table~(\ref{tab:injector_parameter}). The optimization process focuses on normalized emittance, bunch length, and longitudinal phase space. Additionally, energy spread has been incorporated as an optimization constraint. To better align the beam parameters at the module exit with the requirements of the merger section, a further optimization of the beam parameters is conducted based on the initial optimization. This refinement is achieved using Eq.~(\ref{EQ:1}) to enhance the beam parameters.
\begin{equation}
\label{EQ:1}
   max(abs(T_i -A_i)-B_i,0)
\end{equation}
where $T_i$ refers to the optimized target; $A_i$ and $B_i$ is the central value and range of the optimized target. 
\subsection{Merger Design} \label{S2B}

\begin{figure}[!htb]
\includegraphics
  [width=4cm]{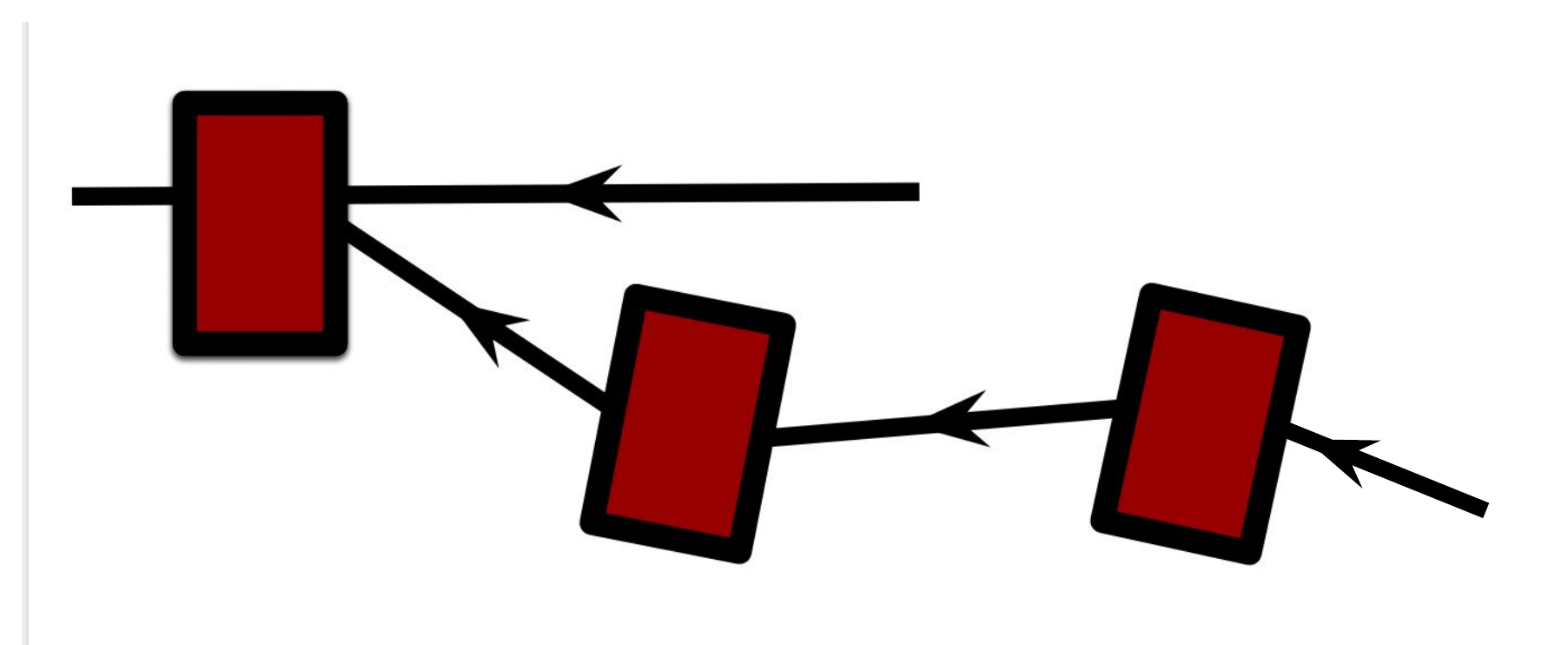} \quad 
  \includegraphics
  [width=4cm]
  {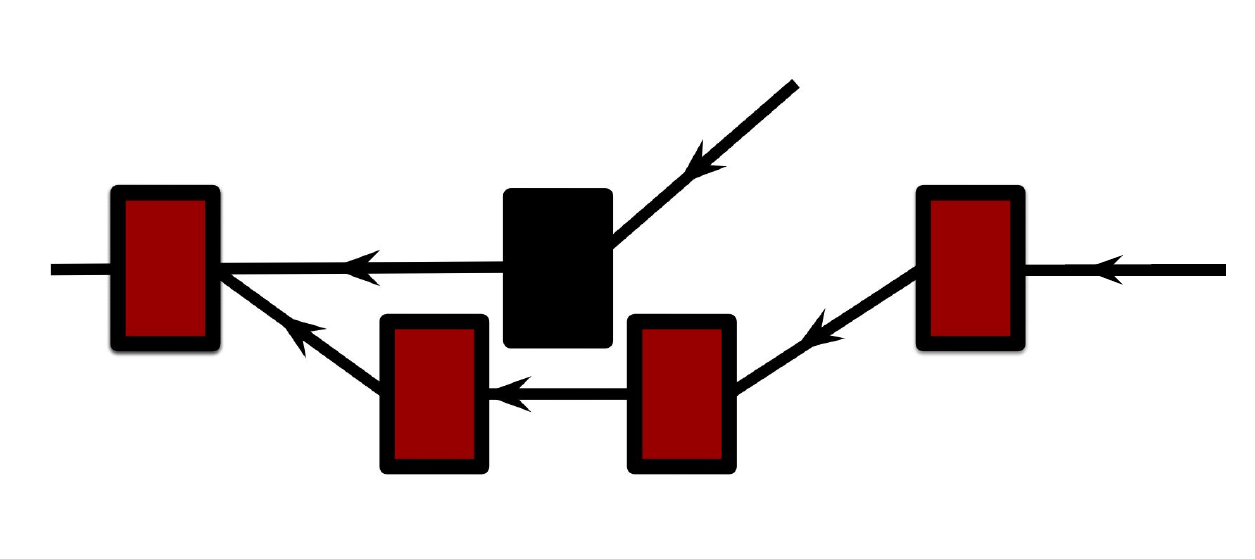}\newline
  (a)\quad\quad\quad\quad\quad\quad\quad\quad\quad\quad\quad(b)\newline
  \includegraphics[width=4cm]{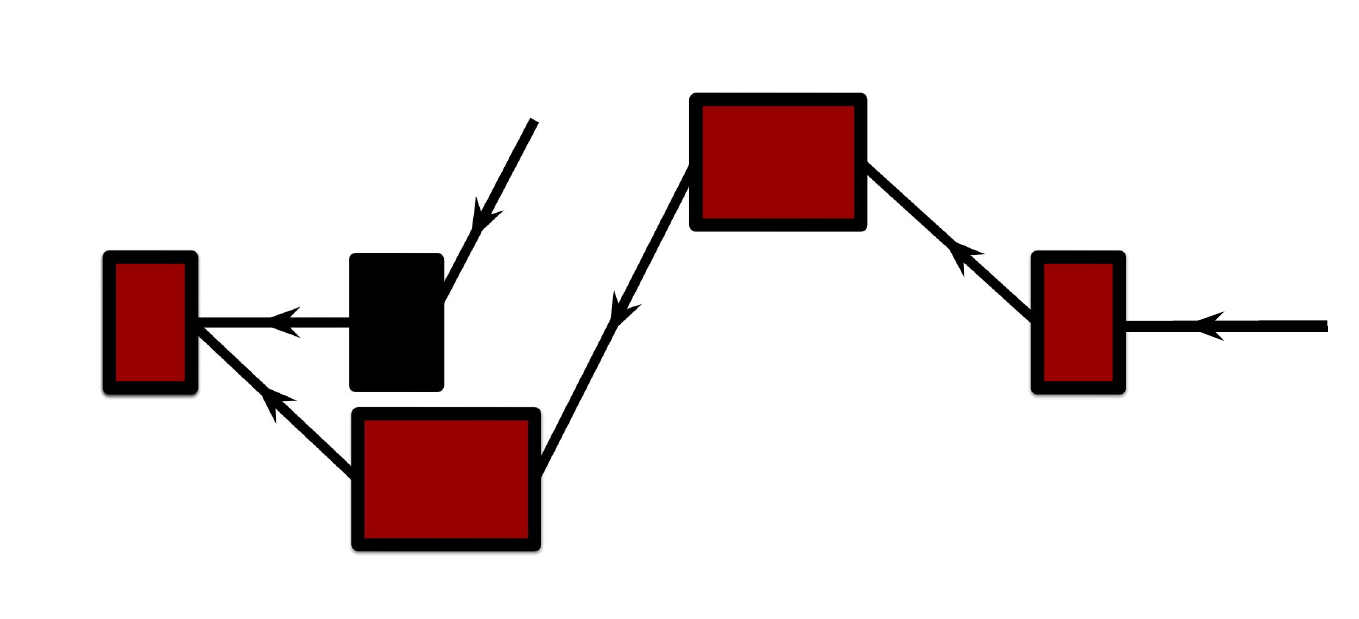}
  \includegraphics[width=4cm]{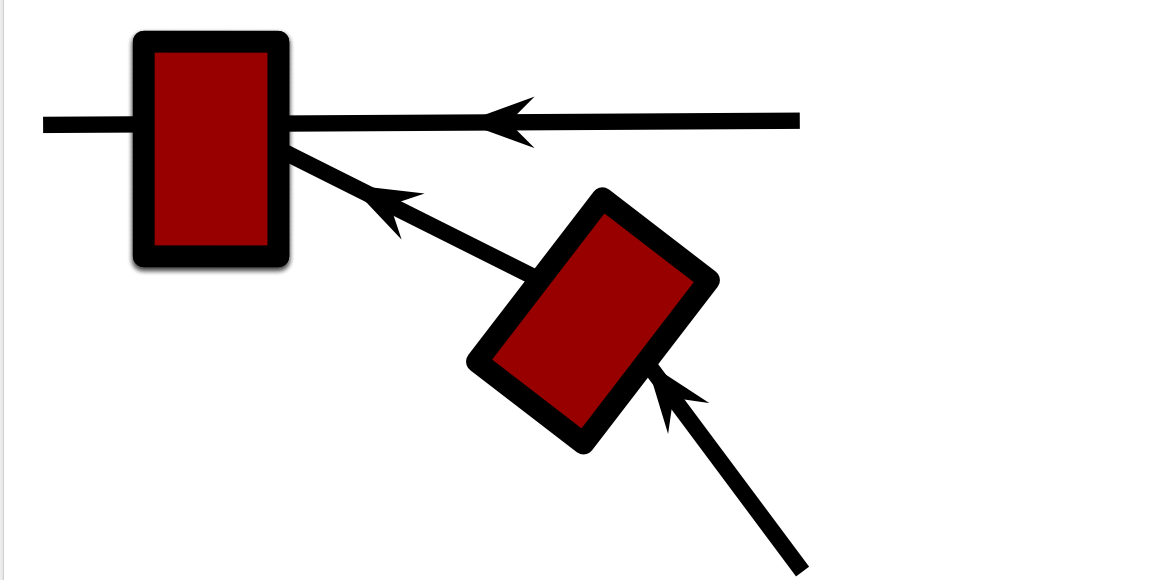}\newline
  (c)\quad\quad\quad\quad\quad\quad\quad\quad\quad\quad\quad(d)
\caption{Common designs for the merger section in ERL include: (a) three bends 3-B merger, (b) chicane merger, (c) zigzag merger, (d) double bend achromat (DBA) merger. Only bend magnets (red and black block) were shown in these figure. In practical designs, it may be necessary to employ quadrupole magnets to make it achromatic.}
\label{fmer1}
\end{figure}
 \subsubsection{Previous Merger Design}
The merger section for ERL machines has been extensively studied\cite{Angal-Kalinin2017, Hoffstaetter2003,Xiong-Wei2009, Sakanaka2014, Gulliford}. Common designs for the merger section were shown in Fig.~(\ref{fmer1}). For high-brightness machines, not only spatial constraints but also emittance preservation are considered to mitigate the impact on the transverse plane caused by the space charge effects. A common approach is to add quadrupole magnets on either side of the middle bend in the three-bends merger (Fig.(\ref{fmer1}) (a)) to make the space charge dispersion close to zero \cite{ Hoffstaetter2003,Hoffstaetter2017,Tanaka,Ohkawa2007}. Especially, a residual dispersion function for single-particle dynamics at the exit of the merger was proposed to compensate for the variance in the x-plane\cite{Bondarenko,Bondarenko2011}. The principal objective of this methodology is to minimize the space charge dispersion, rather than the single-particle dispersion.  Moreover, in most cases, optimizing beam parameters or solving differential equations is a complex and time-consuming process.  As an alternative, a method based on the R matrix has been proposed for calculating dispersion with the LSC effects in the deflection beamline in the space charge regime\cite{Hwang2012}.  The R matrix method has been previously proposed for predicting the impact of steady-state Coherent Synchrotron Radiation (CSR) on the beam\cite{Hajima2004}.  The R matrix method used for predicting LSC effects was expanded to the third order in $s$, where $s$ was the natural coordinate determined by the specific structure of the machines. This is due to the fact that the variance in energy spread caused by LSC varies with $s$ in a nonlinear manner when the length of the bunch in merger is changing. By fitting the variance in energy spread caused by the LSC effect during the transmission of the target bunch through straight sections, and combining this with the R matrix, we can predict the impact of LSC on the transverse plane. Then to suppress the effects of LSC, the mergers need to meet:  1) chromaticity cancellation without collective effects, 2) chromaticity cancellation with collective effects, and 3) sufficient focusing capability both in the $x$ and $y$ planes. 

\subsubsection{Qualitative Theory of Transverse Offset Due to LSC in Merger.}\label{THEORY}

Similar to the effects of CSR in the deflection line, if a kick $\Delta(s)$ is applied to a particle at any point $s$ in the beamline,  The components in the phase space of the particle at point $s$ can be expressed as

\begin{equation}
   x_i(s)=\sum_j R_{ij}^{s_0 \rightarrow s}x_{j,0}+\Delta_i(s).
\end{equation}
The subscript $i=1\sim 6$ denote the phase space components. The subscript "0" indicates the initial value of the corresponding physical quantity. where $R_{ij}^{s_0\rightarrow s}$ are the elements of transfer matrix from $s_0$ to $s$. Then the $x_i$ at the end of beamline is

\begin{equation}
   x_i(s_f)=\sum_{k,j}R_{ik}^{s\rightarrow s_f}R_{kj}^{s_0 \rightarrow s}x_{j,0}+\sum_jR_{ij}^{s\rightarrow s_f} \Delta_j(s).
\end{equation}
Then compared to a particle without a kick, the displacement in $x_i$ is

\begin{equation}
   \Delta x_i(s_f)=\sum_j R_{ij}^{s\rightarrow s_f} \Delta_j(s)
\end{equation}
In the case where only LSC is considered, $\Delta_j\neq 0$ only when $j=6$. (For the TSCF, $\Delta_j\neq 0$ when $j=2,4$). Then the $\Delta_i$ can be replaced as $\frac{d\delta_{LSC}(s)}{ds}$, i.e. the additional energy spread at $s$ (simplified to $\delta_{LSC}^{\prime}(z)$). Furthermore, we assume the particle is located at $z$ in the bunch coordinate. Where $z=0$ is the middle point of bunches.
 When deflection occurs only in the horizontal direction, the transverse displacement caused by the additional energy spread at exit of merger can be expressed as
\begin{equation}
    \label{eqmer2}
    \begin{aligned}
&\Delta x_{_{LSC}}(z)=\delta_{_{LSC}}^{\prime}(s,z) R_{16}^{s_i \rightarrow s_f}\\
&\Delta x^{\prime}_{_{LSC}}(z)=\delta^{\prime}_{_{LSC}}(s,z) R_{26}^{s_i \rightarrow s_f}
\end{aligned}
\end{equation}

\begin{figure}[!htb]
\includegraphics
  [width=0.8\linewidth]
  {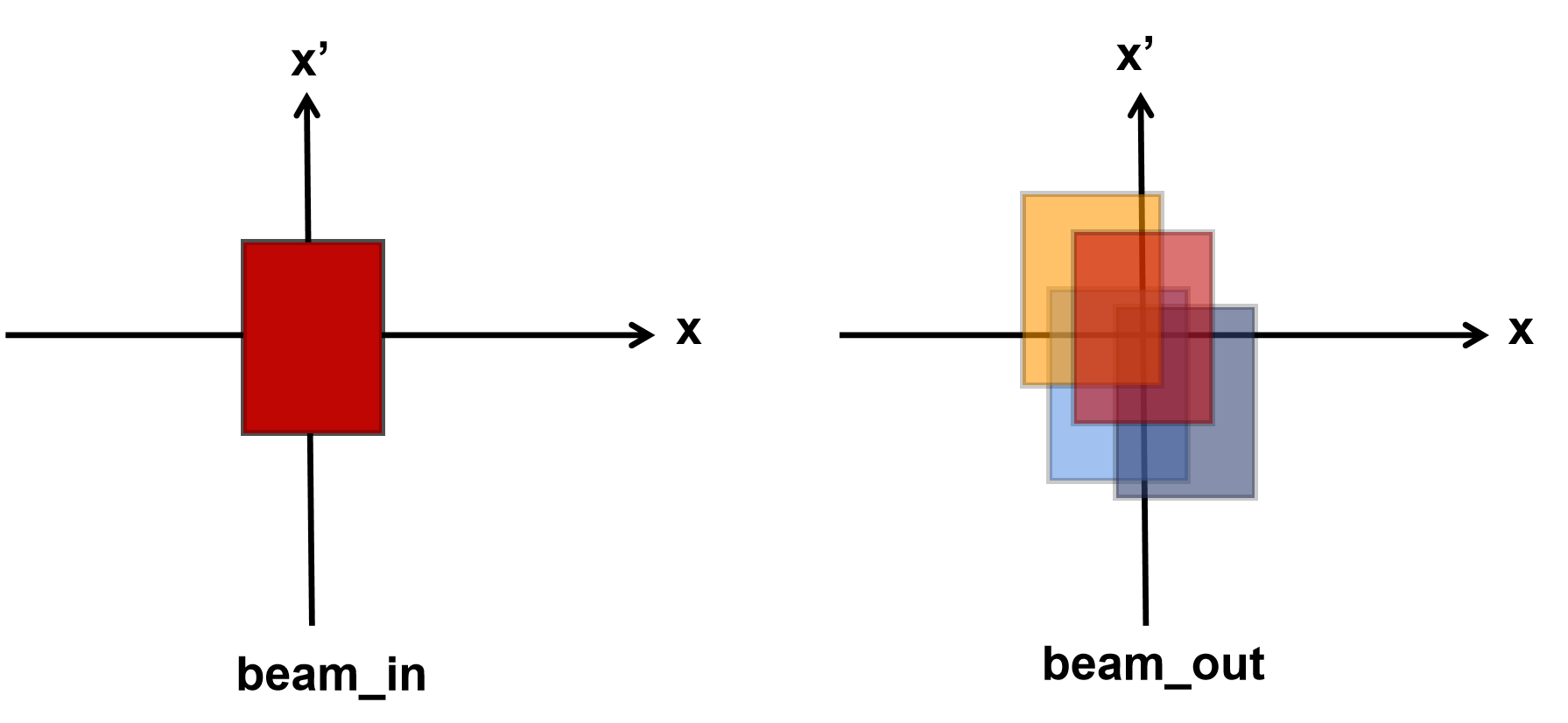} 
\caption{The growth in project emittance caused by the displacement of the slice centers of the bunches.}
\label{fmera1}
\end{figure}

\begin{figure*}[!htb]
\includegraphics
  [width=0.8\linewidth]
  {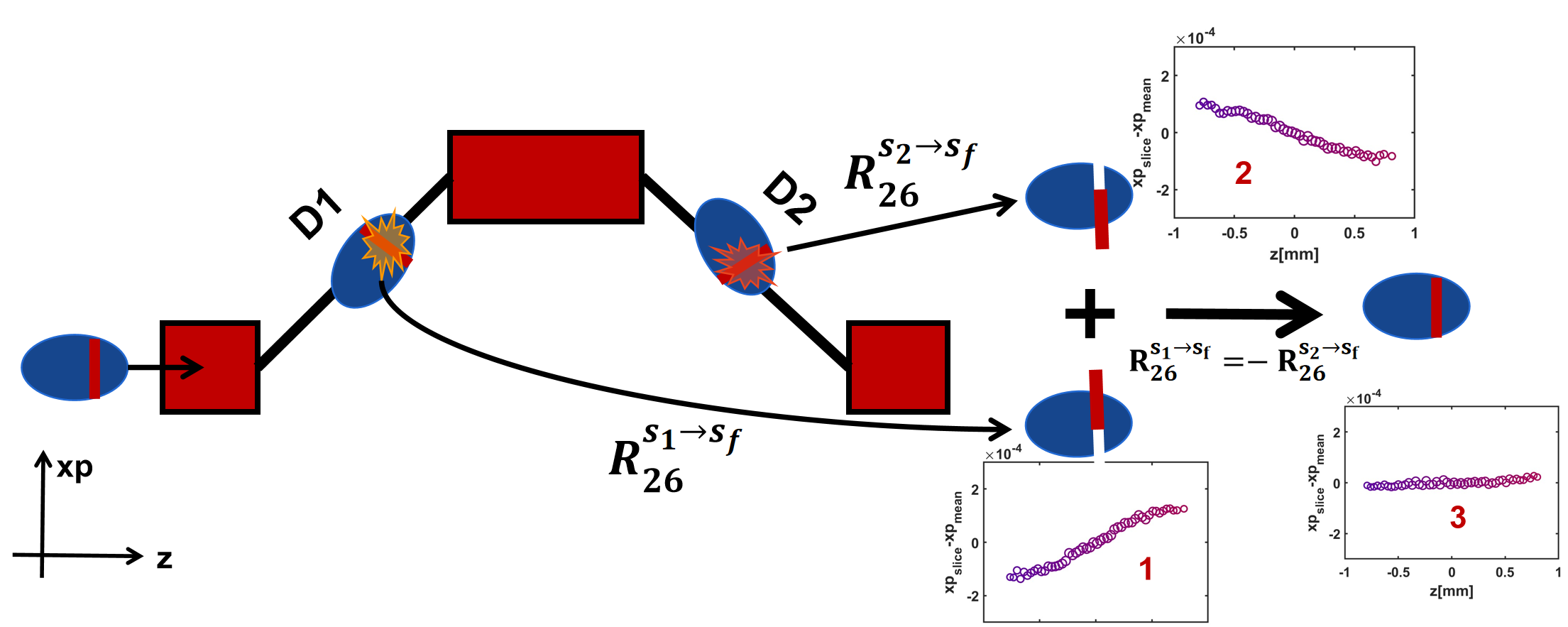} 
\caption{A simple model for suppressing the growth in horizontal emittance. A chicane type, where the length of the second drift is set to zero, was used to simulate and verify the feasibility of this model, which automatically satisfies $R_{26}^{s_1\rightarrow s_f}=-R_{26}^{s_2\rightarrow s_f}$. $s_1$ and $s_2$ the any point in D1 and D2. Inset 1 shows the simulation result where the SC effects occur only in D1. Inset 2: occur only in D2. Inset 3: occur both in D1 and D2. The color of the "o" marker correlates with the $z$, the blue represents the tail, and the red signifies the head. Slices with a slice current less than half the peak current are ignored. }
\label{fmera3}
\end{figure*}
The formula demonstrates that bunch slice centers shift relative to the bunch center in deflection beamlines such as merger and Arc sections, varying for different longitudinal slices and increasing projected emittance as shown in Fig.(\ref{fmera1}). While matching Twiss functions at the exit of beamline can partially mitigate this effect \cite{Hajima2004}, center displacements cannot be fully eliminated due to symplectic properties of matching section matrices. Minimizing ${x}_{_{LSC}}(z)$ for all $z$ is crucial for suppressing projected emittance degradation throughout the machine. Fig.(\ref{fmera3}) illustrates this method: a bunch slice experiencing forces at $s_1$ in D1 and $s_2$ in D2 results in displacements $\delta'(z,s_1)R^{s_1\rightarrow s_f}_{i6}$ and $\delta'(z,s_2)R^{s_2\rightarrow s_f}_{i6}$ respectively at the merger exit. The total displacement, integrated along the path due to distributed LSC effects, can be partially canceled if $R_{i6}^{s\rightarrow s_f}$ has opposite signs at $s_1$ and $s_2$. Simulations comparing LSC effects in D1, D2, and both validate this approach, with Insets 1-3 showing partial cancellation of displacements, resulting in reduced net displacement in $x'$.

 The total transverse offset generated by the LSC effect from the initial point $s_0$ to the observation point $s_f$ is obtained by summing the transverse offsets $\Delta x_{_{LSC}}(z)$ and $\Delta x^{\prime}_{_{LSC}}(z)$ produced at each point along the path. This sum can be represented in the form of an integral, given as
\begin{equation}
    \label{eqmer3}
\begin{aligned}
\hat{x}_{_{LSC}}(z) & =\int_{s_0}^{s_f} \delta_{_{LSC}}^{\prime}(z) \cdot R_{16}^{s \rightarrow s_f} d s \\
\hat{x}_{_{LSC}}^{\prime}(z) & =\int_{s_0}^{s_f} \delta_{_{LSC}}^{\prime}(z) \cdot R_{26}^{s \rightarrow s_f} d s
\end{aligned}
\end{equation}
 This expression is similar to Eqs.(3) and (4) as outlined in Ref\cite{Venturini2015}. However, calculating $R_{16}^{s\rightarrow s_f}$ and $R_{26}^{s\rightarrow s_f}$ at each point and integrating them is complex. Based on the symplectic of the transfer matrix,  we have
 \begin{equation}
M^{-1}=-SM^TS
\end{equation}
$M$ the transfer matrix from $s_0$ to $s_f$. $M^{-1}$ and $M^T$ the reverse  matrix and transpose matrix of $M$. $S$  the standard symplectic matrix. Then we have $R_{16}^{s\rightarrow s_f}=-R_{52}^{s_f\rightarrow s}$ and $R_{26}^{s\rightarrow s_f}=-R_{51}^{s_f\rightarrow s}$, and the Eqs.~(\ref{eqmer3}) can be transformed into the following form
\begin{equation}
    \label{eqmer4}
\begin{aligned}
\hat{x}_{_{LSC}}(z) & =\int_{s_f}^{s_0} \delta_{_{LSC}}^{\prime}(z) \cdot R_{52} d s \\
\hat{x}_{_{LSC}}^{\prime}(z) & =\int_{s_f}^{s_0} \delta_{_{LSC}}^{\prime}(z) \cdot R_{51} d s.
\end{aligned}
\end{equation}
 Where $R_{51}^{s_f\rightarrow s}$ and $R_{52}^{s_f\rightarrow s}$ simplified to $R_{51}$ and $R_{52}$ in this paper. It is important to highlight that the optimization process, following the substitution, proceeds in a reverse direction. This means that the sequence of elements in the optimization is the inverse of their actual arrangement. For instance, the first bend in the optimization process corresponds to the last bend in the actual physical setup. The reverse merger in the optimization process is called as re-merger. 

 In the design work of the merger section, an assumption that can greatly simplify the optimization process is that within the range of the merger, $R_{51}$ and $R_{52}$ satisfy condition $\int_{B} R_{5i} d s <<\int_{all} R_{5i} d s\quad(i=1,2)$. $\int_{B}$ and $\int_{all}$  are the integration ranges over points within the bends and all points in the merger respectively. The $R_{5i}$ in the bends is regarded as constant. Based on the property that $R_{51}$ and $R_{52}$ remain unchanged in the non-bend regions (i.e., the region for the beamline without bends), the Eqs.~(\ref{eqmer4}) can be replaced by

\begin{equation}
\begin{aligned}
\label{reeq1}
\hat{x}_{_{LSC}}(z) &=\sum_i^n R_{52,i}\int\delta'(z)dz=\sum_i^n R_{52,i}\Delta \delta_{_{LSC},i}(z) \\
\hat{x}_{_{LSC}}^{\prime}(z) &=\sum_i^n R_{51,i}\int\delta'(z)dz=\sum_i^n R_{52,i}\Delta \delta_{_{LSC},i}(z)
\end{aligned}
\end{equation}

$n$ is the number of bends for the re-merger. $\Delta \delta_{_{LSC},i}(z)$ is the variance in energy spread between the $i$th bend and $(i+1)$th bends of the re-merger. $R_{51,i}$ and $R_{52,i}$ are the value of $R_{51}$ and $R_{52}$ in the straight sections over the same interval. Furthermore, the variance in energy spread can be expressed in series form:
\begin{equation}
\begin{aligned}
\label{reeq1}
 \delta_{_{LSC},i}(z) =\sum_j^n k_j(z)s^j.
\end{aligned}
\end{equation}
The $k_j(z)$ is the coefficient for the $j$-th term, which can be obtained by fitting the simulation results\cite{Litvinenko2006,Hwang2012}. Substituting into the expression for energy spread, we have 
\begin{equation}
    \label{eqmer5}
\begin{aligned}
\hat{x}_{_{LSC}}(z) &=\sum_{i=1}^{i=n}\sum_{j=1}^{j=m} k_j(z)[(l_{all}-s_{i+1})^j-(l_{all}-s_i)^{j}] \cdot R_{52,i} \\
\hat{x}_{_{LSC}}^{\prime}(z) & =\sum_{i=1}^{i=n}\sum_{j=1}^{j=m} k_j(z)[(l_{all}-s_{i+1})^j-(l_{all}-s_i)^{j}] \cdot R_{51,i}.
\end{aligned}
\end{equation}
 $s_i$ is the position of the $i-$th bend's midpoint in re-merger. $l_{all}$ is the total length for re-merger section. Moreover, the \ref{eqmer5} can be replaced by 
 \begin{equation}
    \label{eq:remer5}
\begin{aligned}
\hat{x}_{_{LSC}}(z) &=\sum_{j=1}^{j=m}k_j(z)\zeta_{sc,j} \\
\hat{x}_{_{LSC}}^{\prime}(z) & =\sum_{j=1}^{j=m}k_j(z)\zeta'_{sc,j} .
\end{aligned},
\end{equation}
where

\begin{equation}
    \label{eqmer6}
\begin{aligned}
\zeta_{sc,j}&=\sum_{i=1}^{i=n}[(l_{all}-s_{i+1})^j-(l_{all}-s_i)^{j})] \cdot R_{52,i} \\
\zeta^{\prime}_{sc,j}&=\sum_{i=1}^{i=n}[(l_{all}-s_{i+1})^j-(l_{all}-s_i)^{j})] \cdot R_{51,i}.
\end{aligned}
\end{equation} 
 The series in Eqs.~\ref{eqmer5} are divided into two parts, as shown in Eqs.~\ref{eq:remer5}: the $k_j(z)$, which is determined by the bunch distribution, and the $\zeta_{sc,j}$ function, defined in Eqs.~\ref{eqmer6}, which is solely related to the lattice parameters.  Therefore, to fully suppress the displacement induced by LSC, the lattice needs satisfy the following condition for any $j$:
 \begin{equation}
    \label{cancellsc}
\begin{aligned}
\zeta_{sc,j}&=0 \\
\zeta^{\prime}_{sc,j}&=0.
\end{aligned}
\end{equation}
These conditions are primarily determined by the lattice characteristics. Mergers that satisfy Eqs.\ref{cancellsc} can effectively suppress LSC-induced displacements for bunches with arbitrary distributions. This implies that the merger sections do not require additional adjustments for varying bunches. In specific cases, high-order $\zeta_{sc}$ functions may be negligible, as will be examined in the following section. Under such circumstances, a simpler merger structure suffices, potentially leading to improved optimization efficiency.


\subsubsection{TBA Merger Design}

\begin{figure}[htpb!]
    \centering
    \includegraphics[width=.6\hsize]{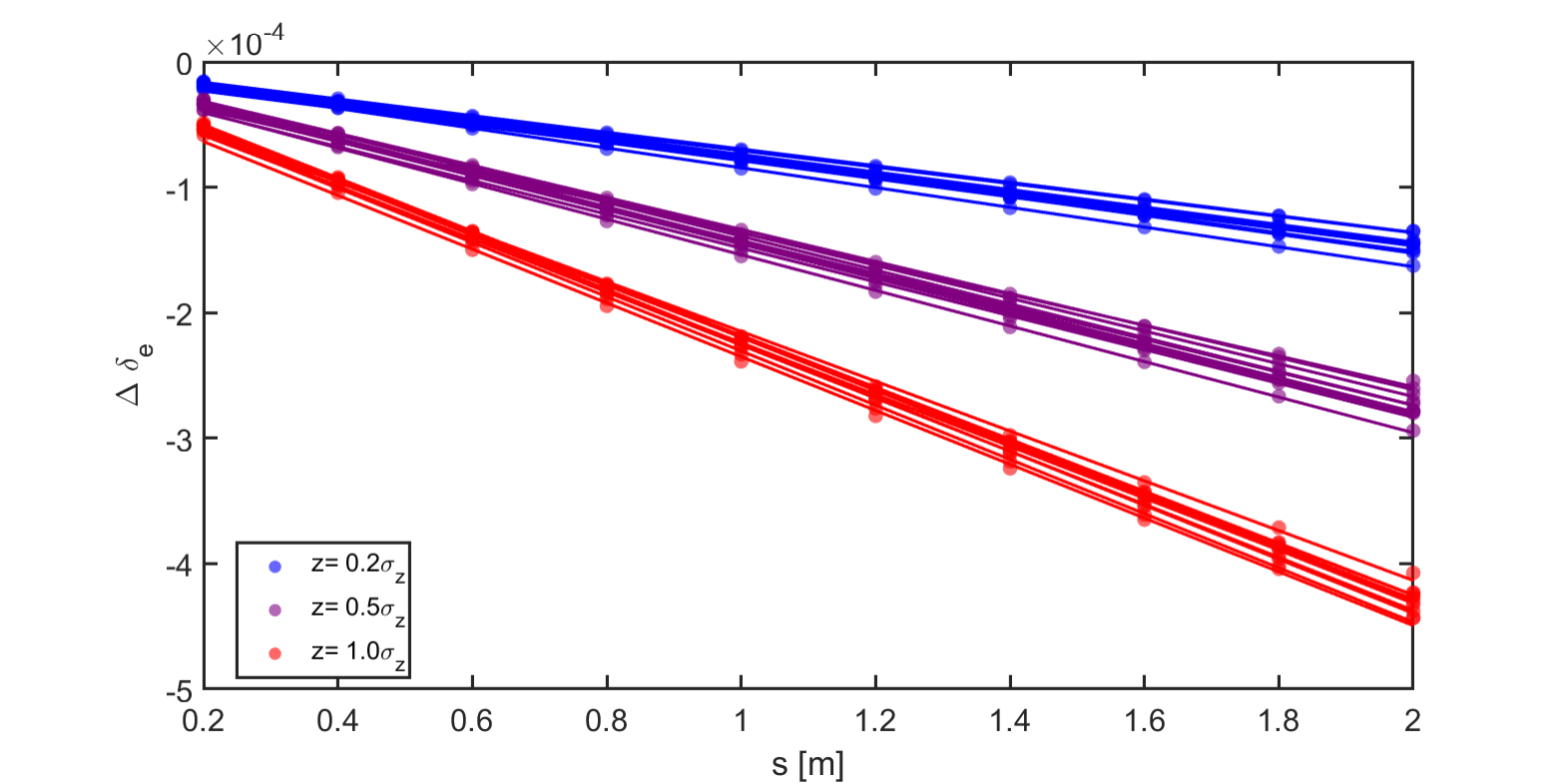}\\
    \includegraphics[width=.6\hsize]{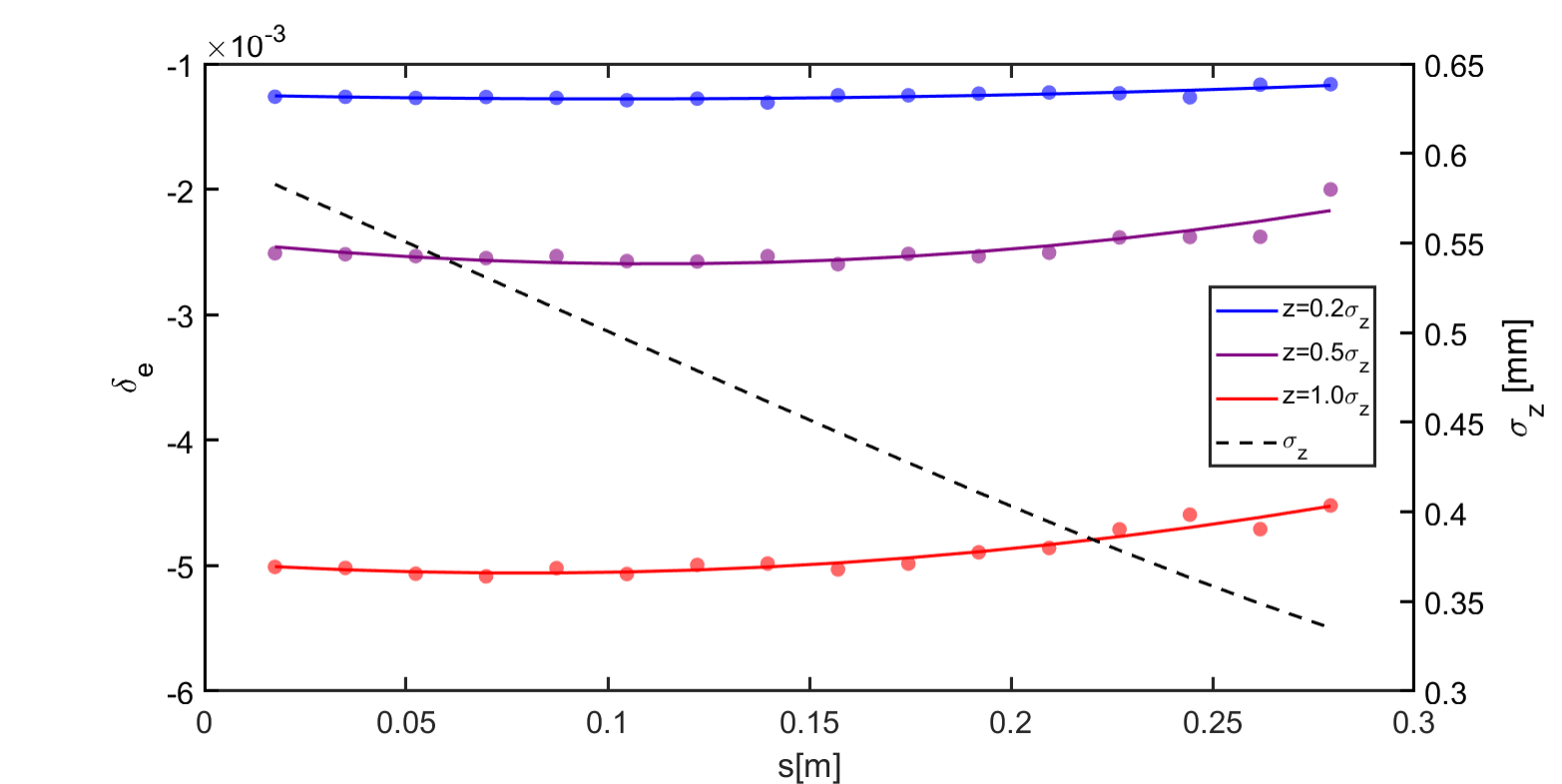}
\caption{The variance in energy spread caused by LSC for three slices: z=$0.2\sigma_z,  0.5\sigma_z, \sigma_z$. The initial bunch length: 0.6mm, the energy: 10MeV, The initial RMS bunch radius: 0.3mm. (Top): the bunch with different transverse momenta in a drift section with $R_{56}$ close to zero. (Bottom): the bunch with a initial energy chirp in a bend with non-zero $R_{56}$. The energy chirp is derived from the simulation results in Sec.\ref{S1A}. The points represent the simulation results obtained from BMad. The curves illustrate the first-order (top) and second-order (bottom) polynomial fits to these data.}
    \label{fig:es}
\end{figure}

In the design work of merger, the $\zeta_{sc}$ function is used to evaluate the effects of LSC in the horizontal direction. And BMad simulation is employed for accelerator simulations\cite{Sagan2006}. 3D space charge methods have been incorporated in the BMad toolkit for charged particle simulations\cite{Mayes, Wang2022}. However, satisfying Eqs.\ref{cancellsc} for all cases of $j$ presents significant challenges. Studies indicate that for a large number of cases, the 2nd order approximate formulas can be used for the expression for $\delta_{_{LSC}}(z)$\cite{Litvinenko2006}:
\begin{equation}
    \label{eqmera2}
    \begin{aligned}
&\delta_{_{LSC}}(s,z)=k_1(z)s+k_2(z)s^2+O(s^3).
\end{aligned}
\end{equation}
 Fig.\ref{fig:es}(top) shows the variance in energy spread for different slices with different transverse evolution. The bunch length is constant for these cases. The variance is approximately linear, which implies that $k_2<<k_1$.Therefore, for the bunches in this study that show no significant changes in bunch length, the first-order $\zeta_{sc}$ function is sufficient. For cases where the variance in bunch length cannot be neglected, the $\delta_{_{LSC}}$ is nonlinear. Fig.~\ref{fig:es}(bottom) shows the simulation results and the second-order fitting curve for different slices.  Thus, for the mergers with large $R_{56}$,  the second-order $\zeta_{sc}$ function must be taken into account.

The specific form of $\zeta_{sc,i}$ for mergers without quadrupole magnets is detailed in \ref{A1}. In these configurations, $\zeta_{sc}$ decreases with reduced drift length after the first bend in the re-merger, with a minimum practical length of 0.6m to avoid beamline intersection. Table \ref{tabmer1} presents the optimal cases for these mergers. The zigzag merger demonstrates superior performance, with its first-order $\zeta_{sc}$ function approaching zero. While effective in suppressing LSC, its compact structure limits practical implementation, especially in high-energy machines where large bends are necessary. Thus, developing a merger that can both suppress LSC effects and be applicable in high-energy machines is crucial for ERLs.

A three-bend merger serves as the optimization baseline, being the simplest structure to achieve a near-zero first-order $\zeta_{sc}$ function. The second-order $\zeta_{sc}$ function incorporates a weight factor $(e^{|R_{56}[m]|}-1)$, with drift section $R_{56}$ neglected. This approach minimizes the impact of the second-order $\zeta_{sc}$ component when the total merger $R_{56}$ approaches zero. The optimization function is then expressed as:
\begin{equation}
    \label{eqmer7}
\begin{aligned}
f_{obj}=C[\zeta_{sc,1}+\zeta'_{sc,1}+(e^{|R_{56}|}-1)(\zeta_{sc,2}+\zeta'_{sc,2})]\quad 
\end{aligned}
\end{equation}

\begin{figure}[!htb]
\includegraphics
  [width=0.8\linewidth]
  {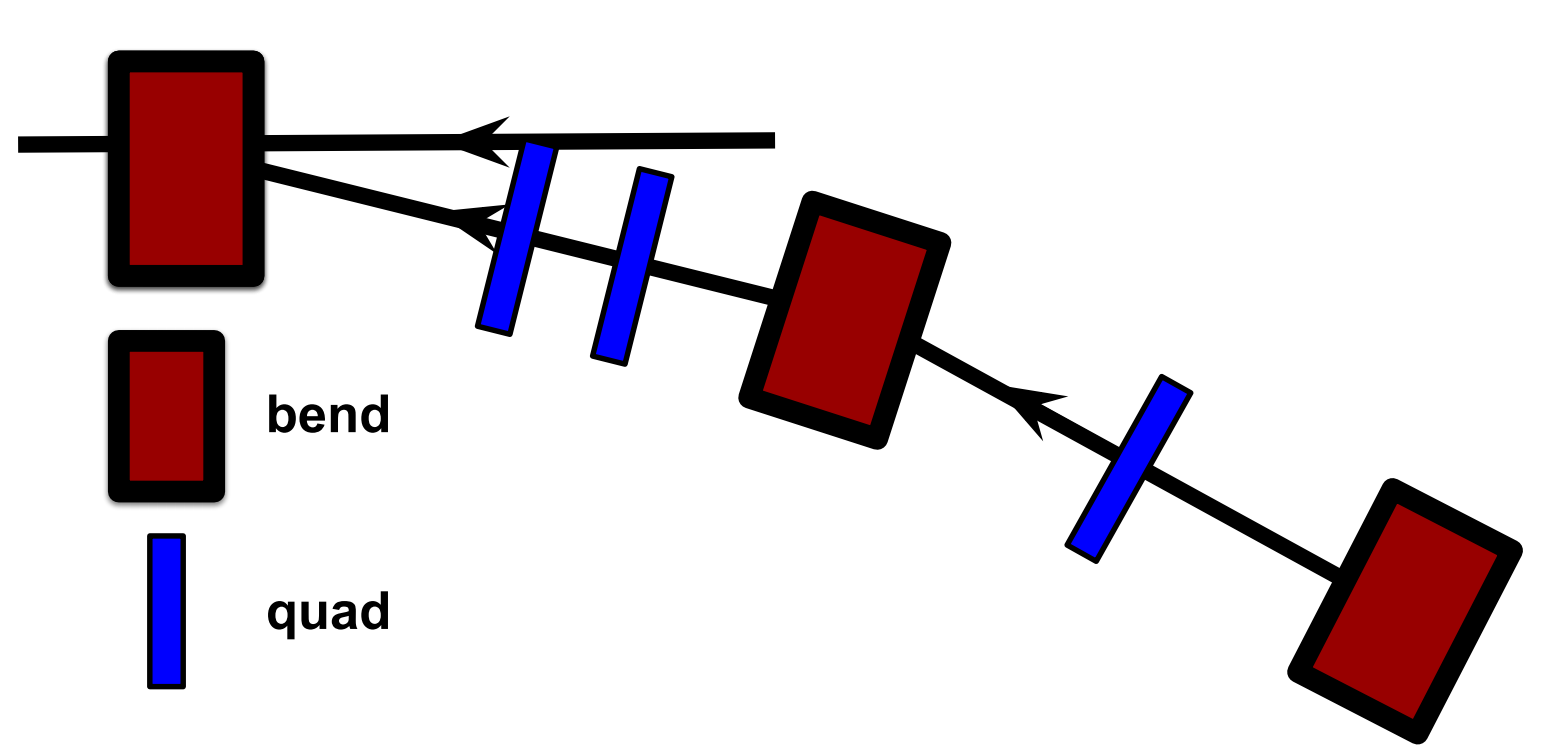}
\caption{The floor plan of the TBA merger.}
\label{fmer2}
\end{figure}

\begin{figure}[!htb]
\includegraphics
  [width=0.8\linewidth]
  {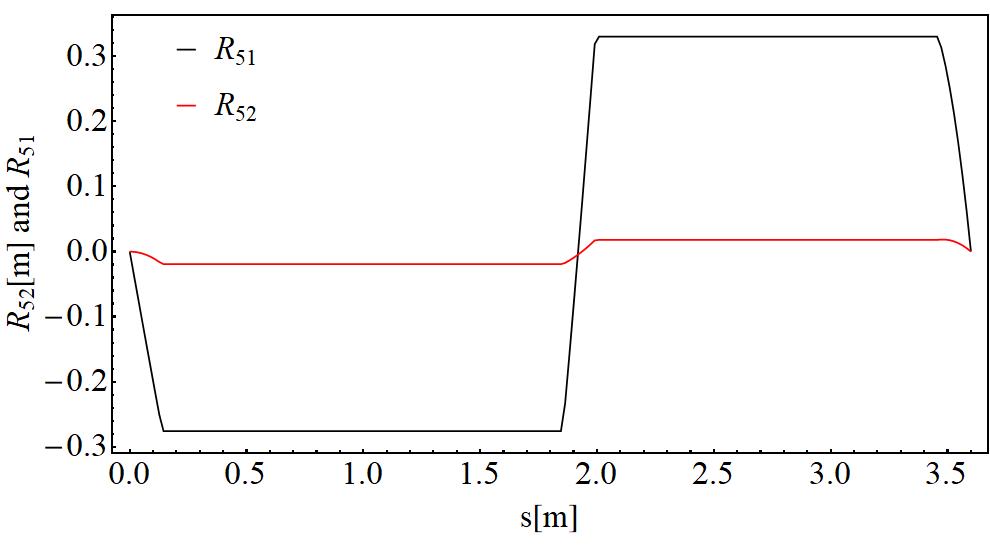}
\caption{The $R_{51}$ and $R_{52}$ of the TBA re-merger.}
\label{fmer3}
\end{figure}

$C$ is the additional factor which is equal to 1 when the merger is achromatic. There are eight variables in this optimization process: three for quad strengths (range $[-30, 30]$), four for drift lengths (range $[0.2, 1]$), and one for the first drift length in the re-merger (range $[0.6, 1]$). The total merger length is constrained to 4m. This study employs the Grey Wolf Optimizer (GWO) \cite{mirjalili2014grey}, an algorithm inspired by grey wolf social hierarchy and hunting behavior, effective for complex optimization problems. The layout of the merger was shown in Fig.~(\ref{fmer2}). It is a TBA with three identical bends. The radius of these bends is 0.5m and the angle is $16^{\circ}$. This structure is referred to as the TBA merger or TBA re-merger in this paper. The $R_{51}$ and $R_{52}$ were shown in Fig.~(\ref{fmer3}). And the $\zeta_{sc}$ functions and $f_{obj}$ for TBA merger were also listed in Table~(\ref{tabmer1}). With its notably smaller values for the first order of $\zeta_{sc}$ functions, the TBA merger demonstrates better suppression of linear LSC effects than other mergers. The second-order $\zeta_{sc}$ function $\zeta'_{sc,2}$ with a higher value leaded to a heightened sensitivity of the TBA merger to the higher-order impacts of LSC. 
\begin{figure}[htpb!]
    \centering
    \includegraphics[width=.45\hsize]{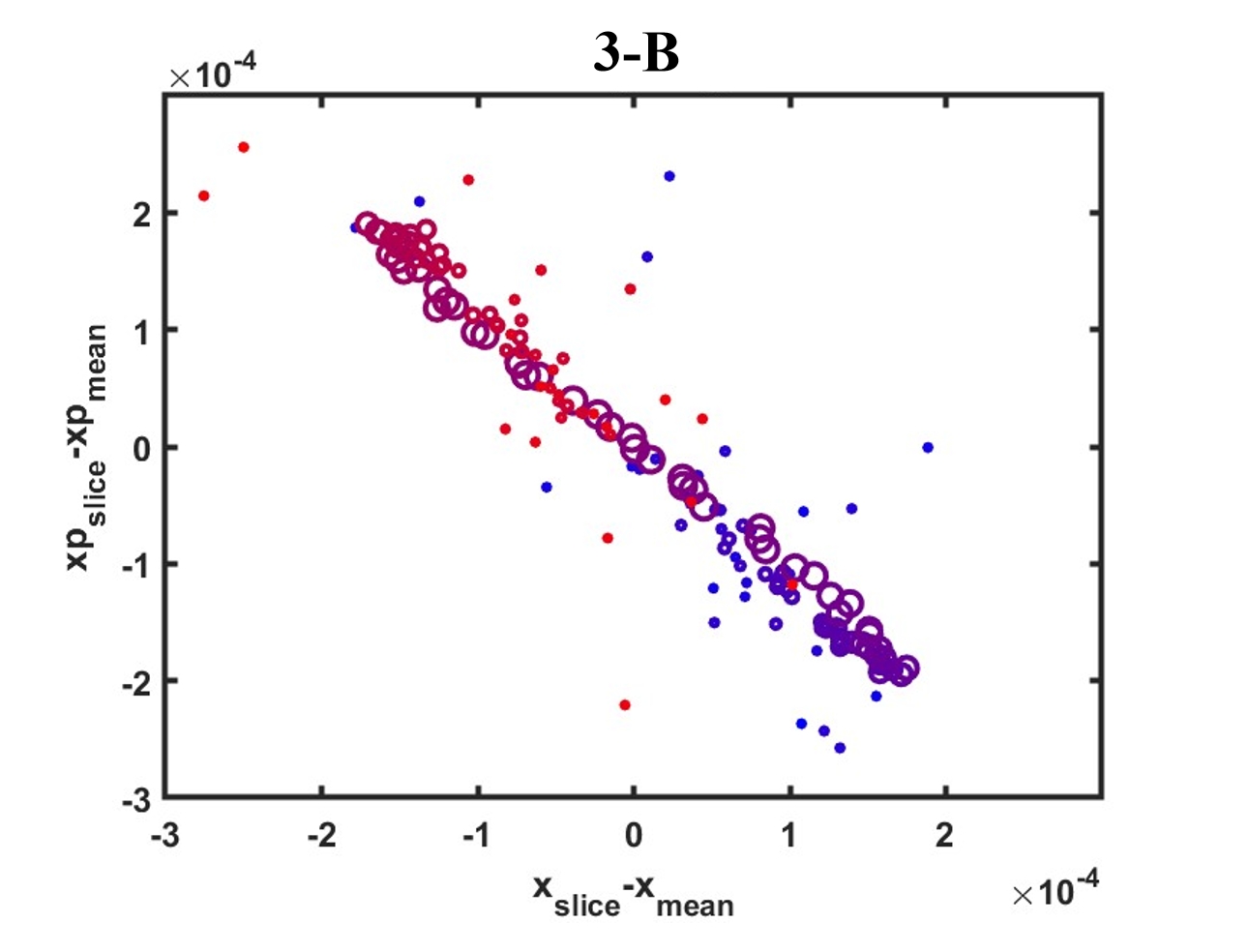}
    \includegraphics[width=.45\hsize]{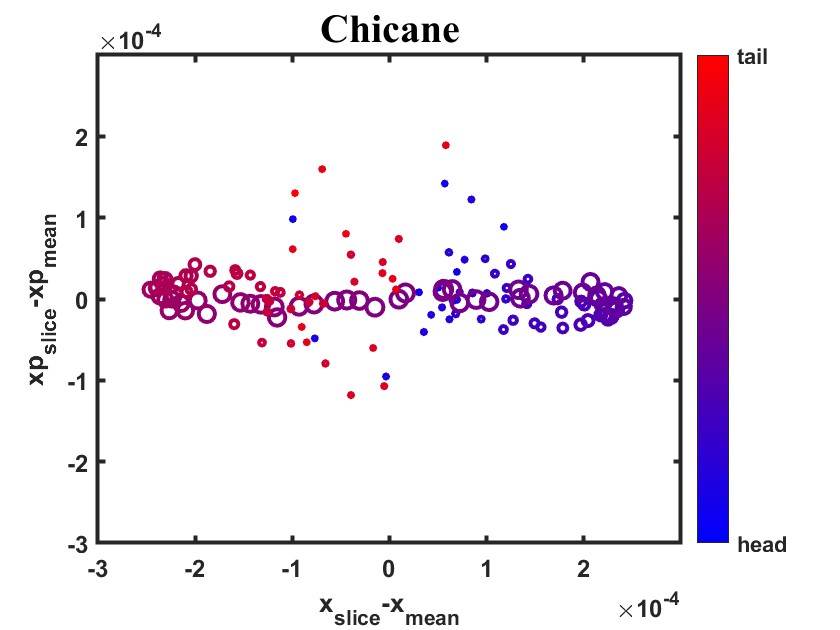}\\
    \includegraphics[width=.45\hsize]{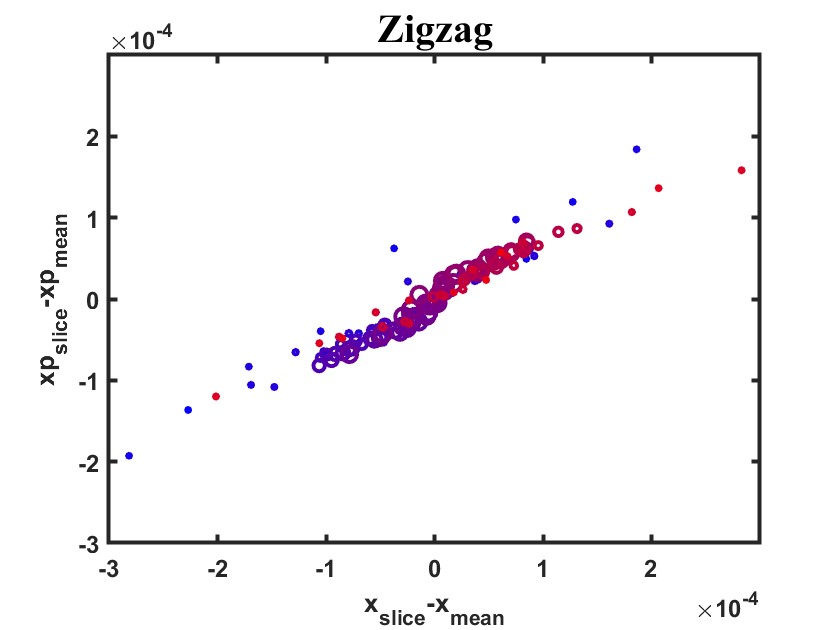}
    \includegraphics[width=.45\hsize]{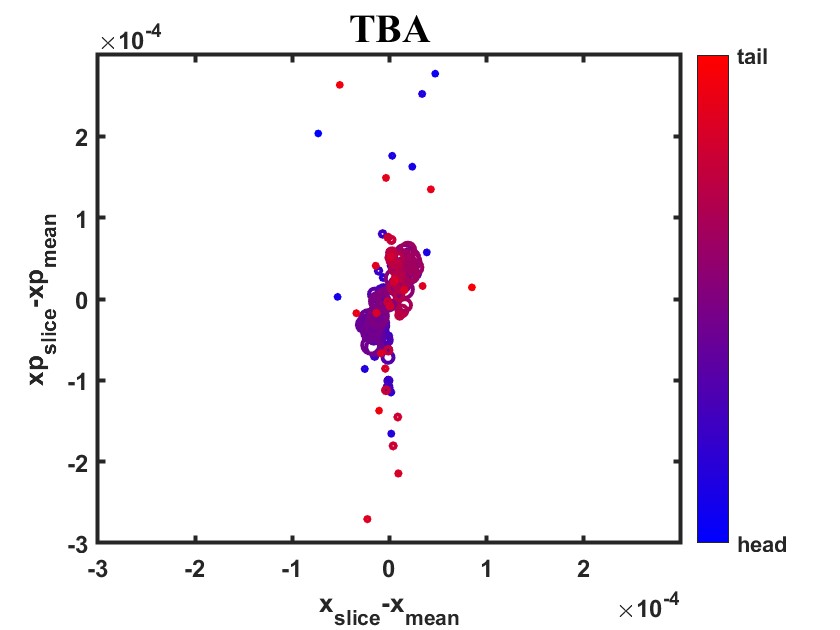}
    \caption{The slice center in the horizontal plane for the bunches from different types of mergers. The size of the 'o' marker correlates with the slice current, and the color corresponds to $z$. The $x_{mean}$ and $x'_{mean}$ are the center of the complete bunch.}
    \label{fmera2}
\end{figure}
However, the smaller $R_{56}$ often results in the bunches being in the "frozen cases," where $k_2$ is significantly lesser compared to $k_1$ in Eqs.~(\ref{eqmera2}). Thus, in most cases, the TBA merger effectively mitigates the influence of LSC in the horizontal plane, even when considering the effects of higher-order terms. 

    \begin{table*}[!htb]
    \centering
\caption{The $\zeta_{sc}$ functions and $f_{obj}$ for different types of merger sections in the best case of \ref{A1}.}
\label{tabmer1}
\begin{tabular*}{16cm} {@{\extracolsep{\fill} } llllllr}
\toprule
Merger type & $\zeta_{sc,1}$ & $\zeta'_{sc,1}$ &$\zeta_{sc,2}$ &$\zeta'_{sc,2}$  &$R_{56}$ &$f_{obj}$\\
\midrule
3-B merger  &0.16 &0.19 &0.08 &0.3 &0.072 &0.37 \\
Chicane merger  & 0.3  & 0 &0.16 &0.2&0.11&0.34 \\
Zigzag merger   &0.015 & 0.023  & 0.63  &0.037&0.08 &0.09 \\
TBA merger & 0.006 & 0.012 & 0.11 &1.4 &-0.002 &0.02 \\
\bottomrule
\end{tabular*}
\end{table*}

\begin{figure}[!htb]
\includegraphics
  [width=0.8\hsize]
  {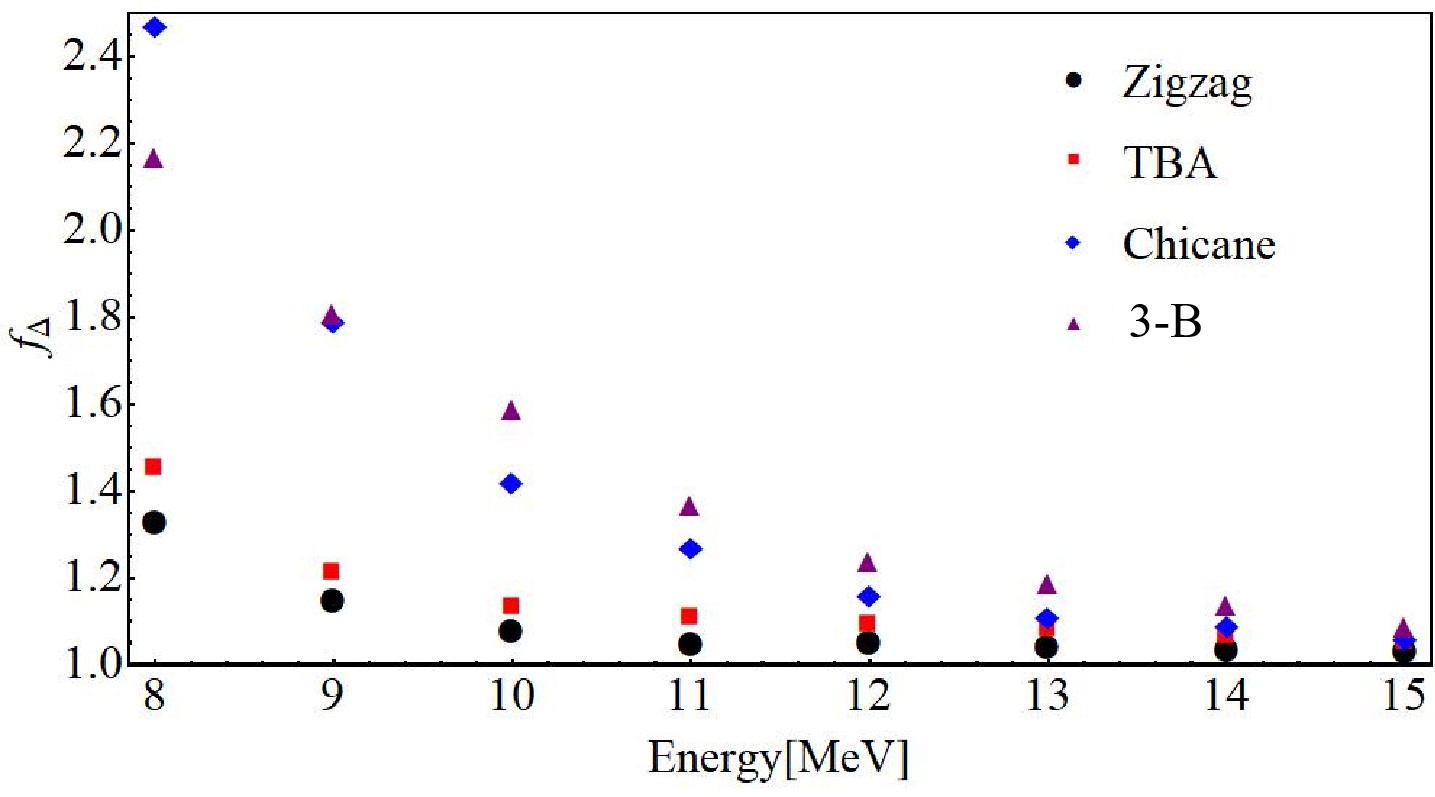}
\caption{Simulation results for different types of mergers with varying energies.}
\label{fmer4}
\end{figure}

Simulations using ideal bunches were conducted to assess efficacy of the TBA merger  in suppressing LSC. The bunch parameters, primarily derived from simulation results of Sec.\ref{S1A} assuming a Gaussian distribution, are presented in Table~(\ref{tabmer2}). The presence of TSCF induces a growth in transverse emittance along the $y$ direction. To evaluate emittance growth in both $x$ and $y$ dimensions concurrently, a function is defined as:
\begin{equation}
    \label{eqmer8}
\begin{aligned}
f_{\Delta}=\mathrm{e}^{(\Delta\varepsilon_x+\Delta\varepsilon_y)/1[\mu\mathrm{mrad}]}
\end{aligned}
\end{equation}
$\Delta \varepsilon=\varepsilon-\varepsilon_0$ represents the emittance variance between  entrance and exit of mergers. Twiss functions at the entrance are optimized to minimize $f_{\Delta}$. Fig.(\ref{fmer4}) illustrates simulation results for various merger types. Under identical initial Twiss functions, space charge-induced emittance growth decreases with decreasing energy. For energies around 15 MeV, LSC in merger sections becomes negligible for the bunch specified in Table(\ref{tabmer2}). Between 8-15 MeV, the TBA merger significantly outperforms 3-B and chicane mergers, but is less effective than the zigzag merger in space charge suppression. This is due to the optimization process being focused solely on LSC effects in the x-direction, failing to provide effective focusing in the y-direction. However, adding additional quadrupole magnets would increase the overall length of the merger section, which is detrimental for slice emittance preservation due to inhomogeneity of energy change in slices.\cite{Bondarenko2011}. Therefore, we have decided to retain the current design after evaluation for the TBA mergers with different numbers of quadrupole magnets. Furthermore, as the discussion in Sec.\ref{THEORY}, the variance in projected emittance in merger sections arises from different slices having different centers.  Simulations were conducted with bunch energies set to 10 MeV and initial Twiss functions matching those in Fig.(\ref{fmer4}). Fig.(\ref{fmera2}) illustrates the slice centers for various merger types. Upon excluding slices with insufficient macroparticles, the TBA merger demonstrates the lowest overall eccentricity in slice center distribution. Smaller $\zeta_{sc,1}$ functions correspond to reduced offsets, as evident in the zigzag and TBA mergers, and in the $x'$ direction of the chicane merger. However, due to the nonlinearity of $\Delta \delta_{_{LSC}}$, these reductions do not correlate linearly with $\zeta_{sc,1}$ function values. The influence of $\zeta_{sc,2}$ functions is manifest in the nonlinear distribution of slice centers in zigzag and TBA mergers. The larger $\zeta'_{sc,2}$ of TBA merger results in greater slice center displacements along the $x'$ direction. Moreover, displacements show significant correlations with longitudinal positions. Bunch heads and tails shift in opposite directions in phase space, attributable to the LSC process where $\delta'_{{LSC}}$ signs differ for head and tail. Consequently, the bunch head gains energy while the tail loses energy.

\begin{table}[!htb]

\centering
\caption{Parameters of the electron beams.}
\label{tabmer2}
\begin{tabular*}{8cm} {@{\extracolsep{\fill} } lllr}
\toprule
Parameter & Value &     Units \\
\midrule
Bunch charge & 100 &pC\\
Bunch length & 0.6 &mm\\
Energy & 8-15 & MeV\\
Slice energy & 0.01 &$\%$\\
$\varepsilon_{x_0\&y_0}$ & 1 &$\mu$mrad\\
\bottomrule
\end{tabular*}
\end{table}

\begin{figure}[!htb]
\includegraphics
  [width=0.8\hsize]
  {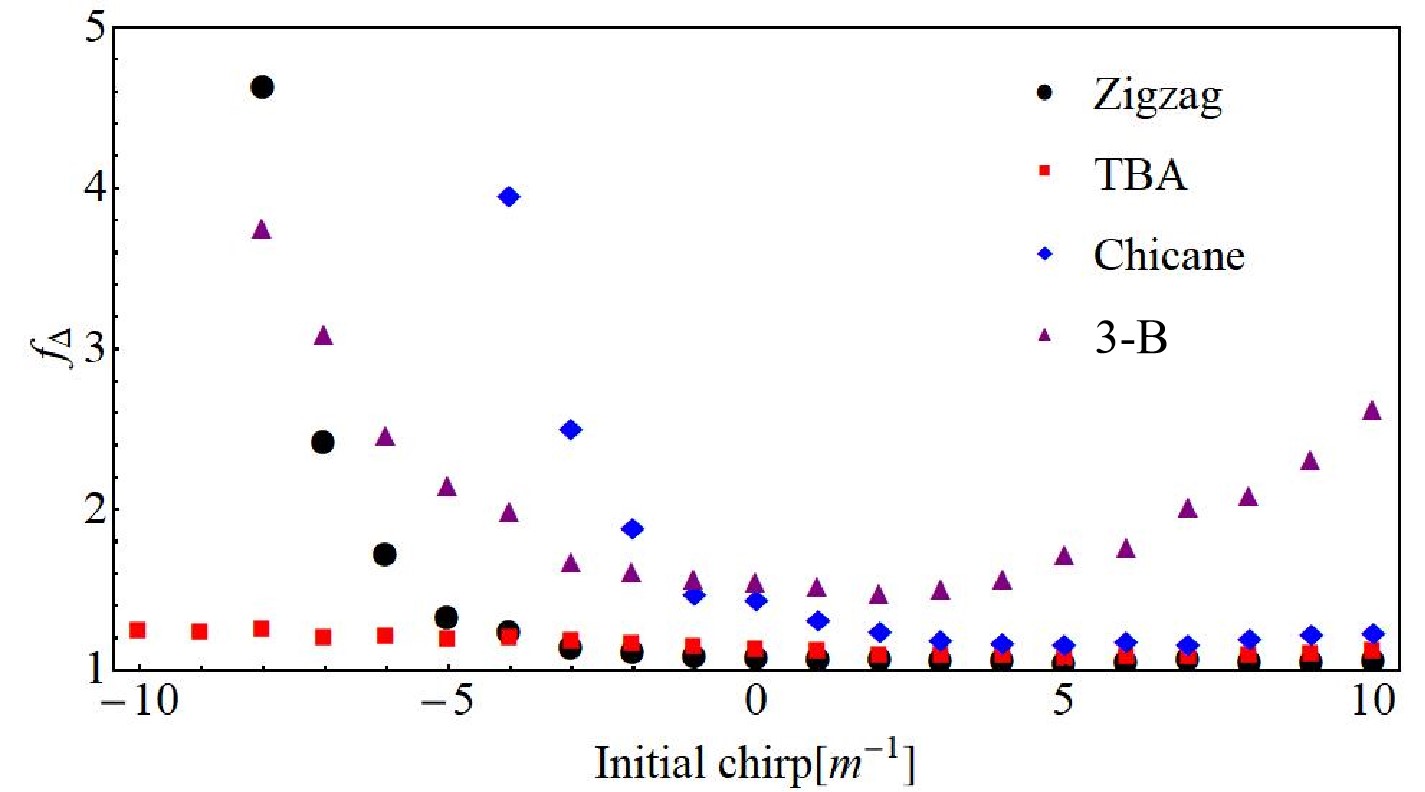}
\caption{Simulation results for different types of mergers with different initial energy chirps.}
\label{fmer5}
\end{figure}

In practical applications, energy chirp is an unavoidable phenomenon for bunches. This chirp primarily stems from two sources: the energy modulation induced by space charge effects in the injector, and the impact of the accelerating cavity's phase. Excessive chirp, when coupled with the $R_{56}$ of the merger, can lead to longitudinal bunch distortion, potentially compromising transmission efficiency in subsequent sections. Consequently, a greater tolerance for energy chirp provides a wider optimization range for the accelerating phase during the injection optimization process. A series of bunches with varying initial chirps were simulated to evaluate the merger's capacity to suppress LSC effects. The initial chirp was scanned within the range of -10 to 10, with the bunch energy set to 10 MeV. The simulation results are presented in Fig.~(\ref{fmer5}). From the simulation results, it is evident that the emittance of the TBA merger was well preserved across the entire range of initial chirps. When the initial chirp is less than zero, other types of merger experience an increase in emittance. This growth in emittance is primarily due to the significant positive $R_{56}$ values of these mergers couple with initial chirp leaded to a decrease in bunch length. The compressed bunches exhibit increased density, enhancing their susceptibility to LSC effects. Simultaneously, this compression makes $k_2$ in Eqs.\ref{eqmer2} non-negligible, reducing the performance of mergers. To mitigate these issues, the accelerating phase of the injector cavity is adjusted and constrained during optimization. However, this phase constraint inherently limits the available solution space for optimization. But it also leads to a different challenge: the particle bunches tend to become elongated or stretched. An elongated bunch is more challenging to compress effectively in subsequent stages of the beam compressor process\cite{DiMitri2016}.

\section{Start to End simulation for H-B injector} \label{S3}
\begin{figure}[!htb]
    \centering
    \includegraphics[width=\linewidth]{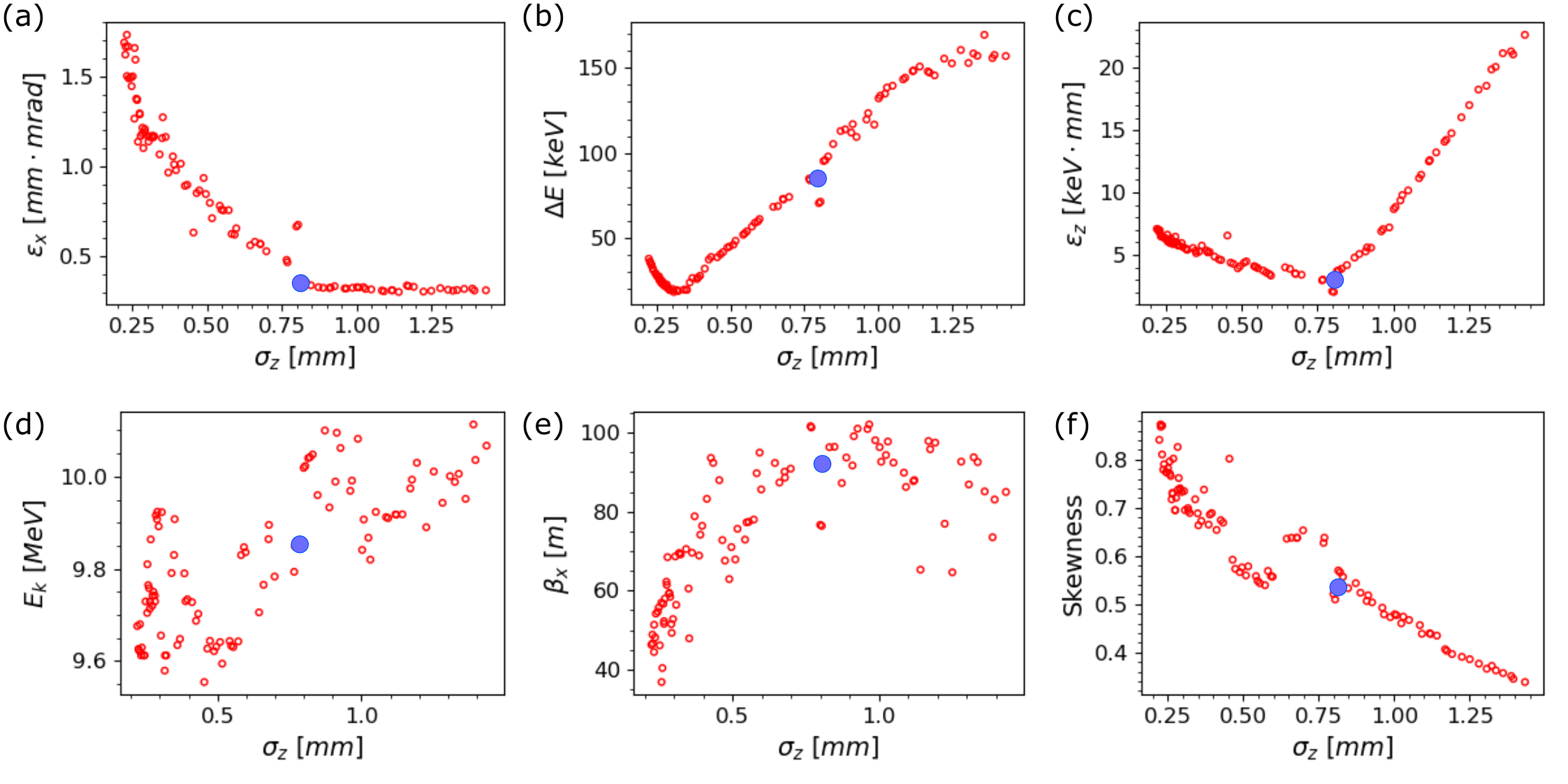}
    \caption{The figure represents the correlations of multi-objective optimization for the export beam parameters of the module. The horizontal axes represent the bunch length of the beam, and the vertical axes represents the emittance, energy spread, longitudinal emittance, kenitic energy, $\beta_x$ and skewness of longitudinal profile respectively.  The chosen parameters for the beam are the emittance of 0.4 $mm\cdot mrad$, the bunch length of 0.81 mm, the $\epsilon_z$ of 3.7 $keV\cdot mm$, the $\beta_x$ of 90 m, the skewness of 0.55.}
    \label{fig:injet_opt}
\end{figure}

\begin{figure}[htbp!]
    \centering
    \subfigure{\includegraphics*[width=\linewidth]{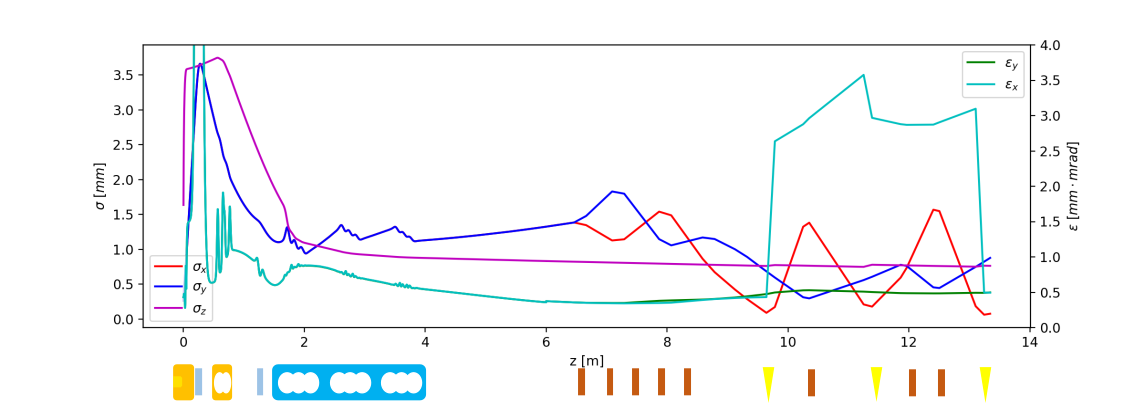}}
    \caption{The figure contains the evolution of transverse normalized emittance, bunch length, beam size. The normalized emittance at the exit of merger is 0.5$mm\cdot mrad$, the beam size ($\sigma_x$ and $\sigma_y$) is 0.3 mm and 1.5mm. The bunch length is 0.77 mm.}
    \label{fig:evo_beam}
\end{figure}

\begin{figure}[!htb]
    \centering
    \includegraphics[width=\linewidth]{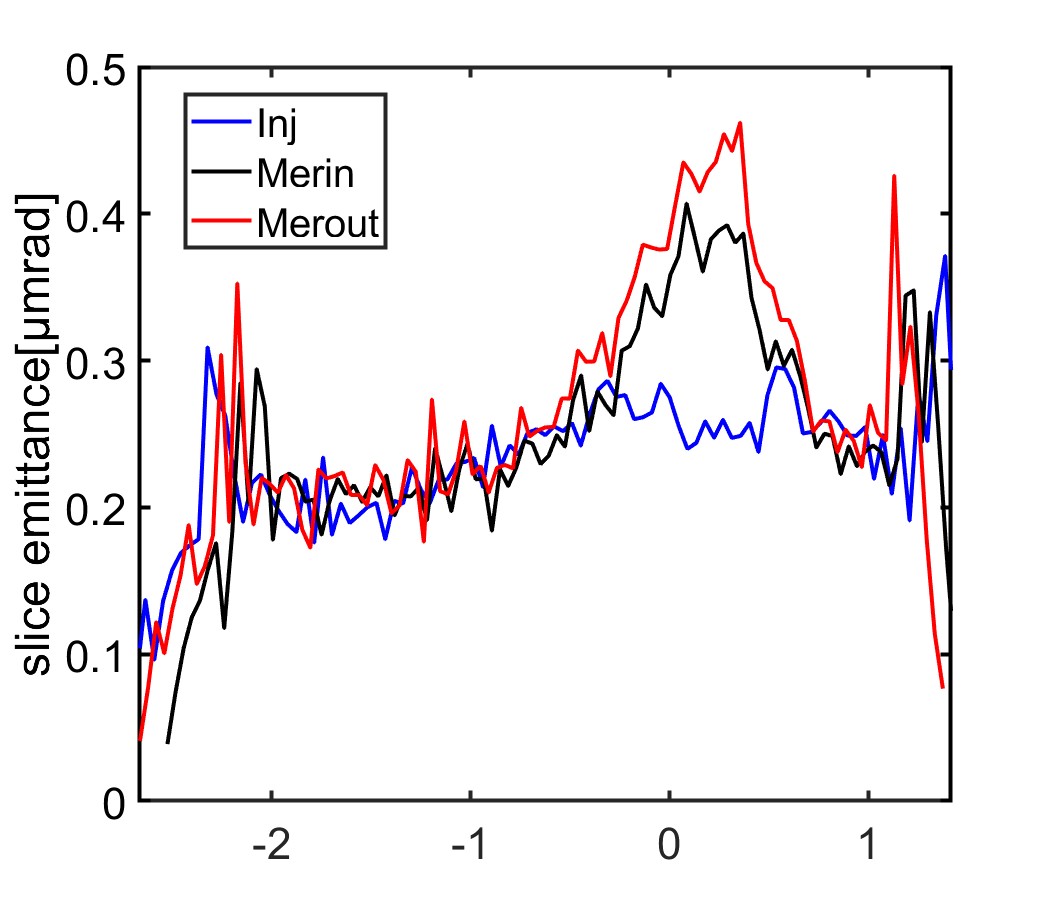}
    \caption{The slice emittance in horizontal plane at different positions of the beamline.}
    \label{sli_emit}
\end{figure}

The optimization of the injector primarily involved the use of ASTRA and BMad  for optimizing the injector and merger section, respectively. The objective is expressed as $a = \sigma_z,b = \epsilon + I_s/10 + \sigma_{HO}/10$. $I_s$ represents the skewness of the longitudinal current distribution of the beam. It can be calculated using formula $E[(\frac{z-\mu}{\sigma_z})^3]$, where $\mu$  is the mean  of the longitudinal distribution. $E$ denotes the mean operation. In the injection section, the optimization focused on the cavity phase, position, and field strength, as well as the parameters of the solenoid. The optimized parameters of beam are shown in Fig. ~(\ref{fig:injet_opt}), and the blue dot indicates the operating point of the injector selected in this study. In the merger section, the optimization mainly involved the quadrupole magnets (beam matching) and dipole magnets. The optimization function primarily includes beam emittance, bunch length, higher-order energy spread, energy spread, and current skewness.
The main parameters of injector's components were shown in Table~(\ref{tab:injector_parameter}), and the evolution of transverse normalized emittance, bunch length and beam size was shown in Fig. ~(\ref{fig:evo_beam}). The bunch parameters were listed in Table~(\ref{tab:bunch_para}). The bunches are derived from the simulation results obtained at (a) the exit of the injector, (b) the exit of the match section between the injector and the merger, and (c) the exit of the merger. Compared to the bunches at the exit of the injector, the variance in horizontal or vertical emittance at the exit of the merger was controlled to within $0.1$. A significant growth in slice emittance occurs in the slices with large current after a bunch compression process in match section. This is mainly attributed to the increase in slice emittance, as illustrated in Fig.~(\ref{sli_emit}). This variance is primarily due to the nonlinear space charge force. The typical beam distribution in the transverse direction is Gaussian-like, where $R/\sigma_{x,y}$ is greater than 2. This can lead to strong nonlinear space charge forces in any transverse distribution.\cite{PhysRevSTAB.15.090701,PhysRevAccelBeams.24.124402}. $R$ represents the beam radius, while $\sigma_{x,y}$ denotes the rms in the transverse direction. Additionally, due to the axially symmetric field distribution before the match section, we have $\sigma_x=\sigma_y$. Then the slice emittance at different positions along the beamline was shown in Fig.~(\ref{sli_emit}). A cutoff Gaussian distribution can effectively reduce the nonlinear space charge force by decreasing the value of $R/\sigma_{x,y}$\cite{PhysRevAccelBeams.24.124402}. The change in slice emittance can also be well controlled. However, for ERLs, it is not advisable to truncate the bunch lightly. After evaluation, we have chosen to retain the existing case.

\begin{table}
\centering
\caption{Parameters of the injector components.}
\label{tab:injector_parameter}
\begin{tabular*}{8cm} {@{\extracolsep{\fill} } llllr}
\toprule
Component & Max. field &  Phase & Position/units \\
\midrule
Gun  & 20 MV/m & -3 & 0
\\
Sol. 01  & 0.0613 T & ~ & 0.25
\\
Sol. 02  & 0.0292 T & ~ & 1.27
\\
Buncher  & 2 MV/m & -69.5 & 0.67
\\
CAV01  & 18 MV/m & -30.0 & 1.8
\\
CAV02  & 18.5 MV/m & -23.3 & 2.7
\\
CAV03  & 21.2 MV/m & -15.0 & 3.6
\\
Quadruple (merger) $k$   &\multicolumn{2}{l}{27.7,-14.1,26.47} & ~
\\
 Dipole (merger)    &\multicolumn{2}{l}{16/0.5} & deg/m
 \\
\multirow{2}*{Laser}   &Duration & 42.2 &ps
\\
~                      &Diameter &0.189  &mm
\\
\bottomrule
\end{tabular*}
\end{table}

\begin{table}[!htb]
\centering
\caption{Bunch parameters of the S2E simulations. "Inj" the bunch at the exit of injector. "Merin" the bunch at the entrance of the Merger. "Merout" the bunch at the end of merger}
\label{tab:bunch_para}
\begin{tabular*}{8cm} {@{\extracolsep{\fill} } llllr}
\toprule
Parameter[Units] & Inj &  Merin & Merout \\
\midrule
Bunch Charge[pC] & 100 & 100 &100\\
Energy[MeV]  & 10.7 & 10.7 & 10.7
\\
Bunch Length[mm] & 0.83 & 0.77 & 0.75
\\
Peak Current[A] &  16 &  18 & 18 \\
Slice Energy Spread[$\%$] &0.02 & 0.02 &0.02\\
Energy Spread[$\%$] &0.93 & 0.89 &0.87\\
Emittance (x/y)[$\mathrm{\mu mrad}$] & $0.38/0.38$    &$0.43/0.49$ & $ 0.47/0.47$\\
\bottomrule
\end{tabular*}
\end{table}

\begin{figure}
    \centering
    \includegraphics[width=.45\hsize]{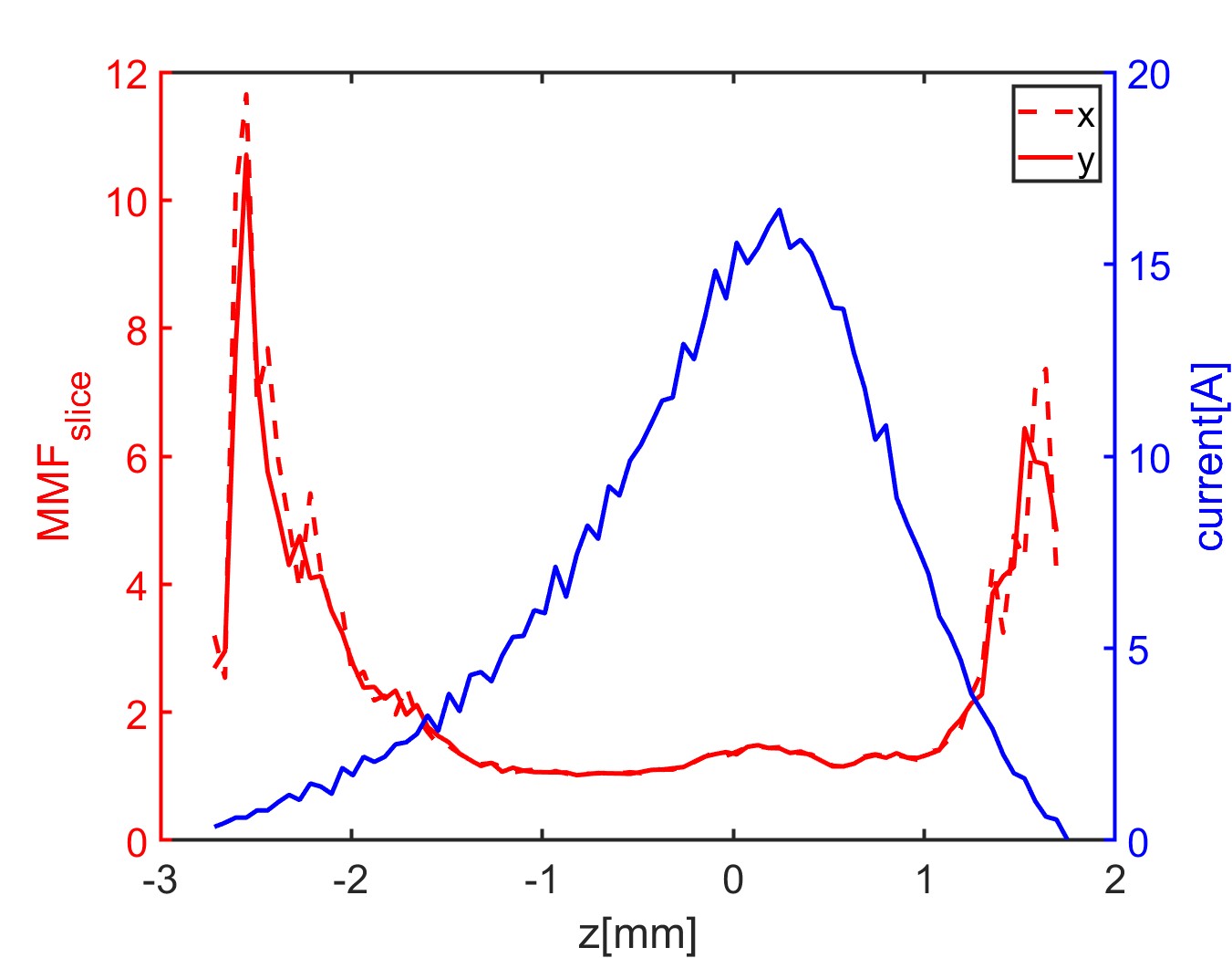}
    \includegraphics[width=.45\hsize]{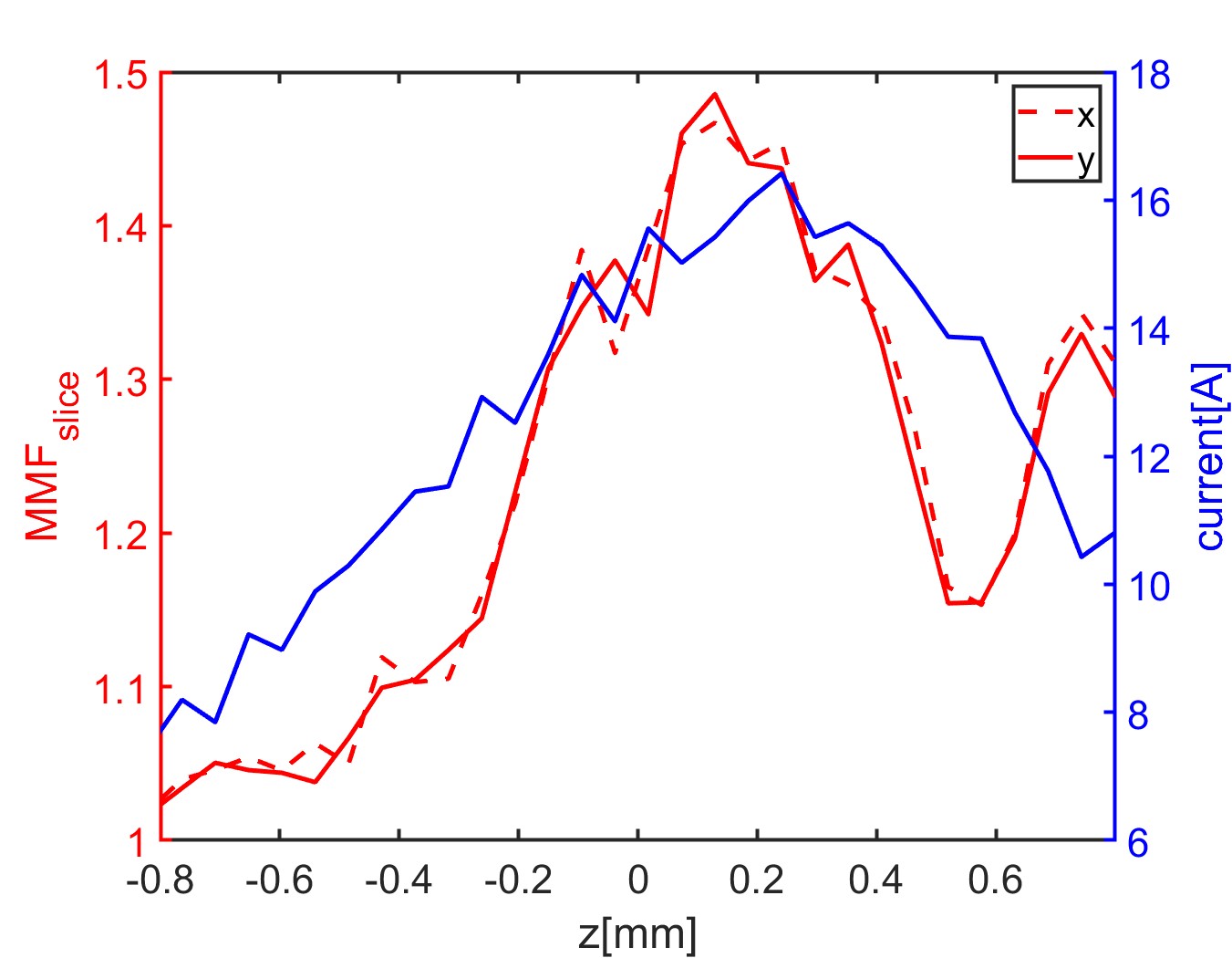}
    \\
    1(a) \hspace{3.2cm} 1(b)\hspace{4cm}\\
        \includegraphics[width=.45\hsize]{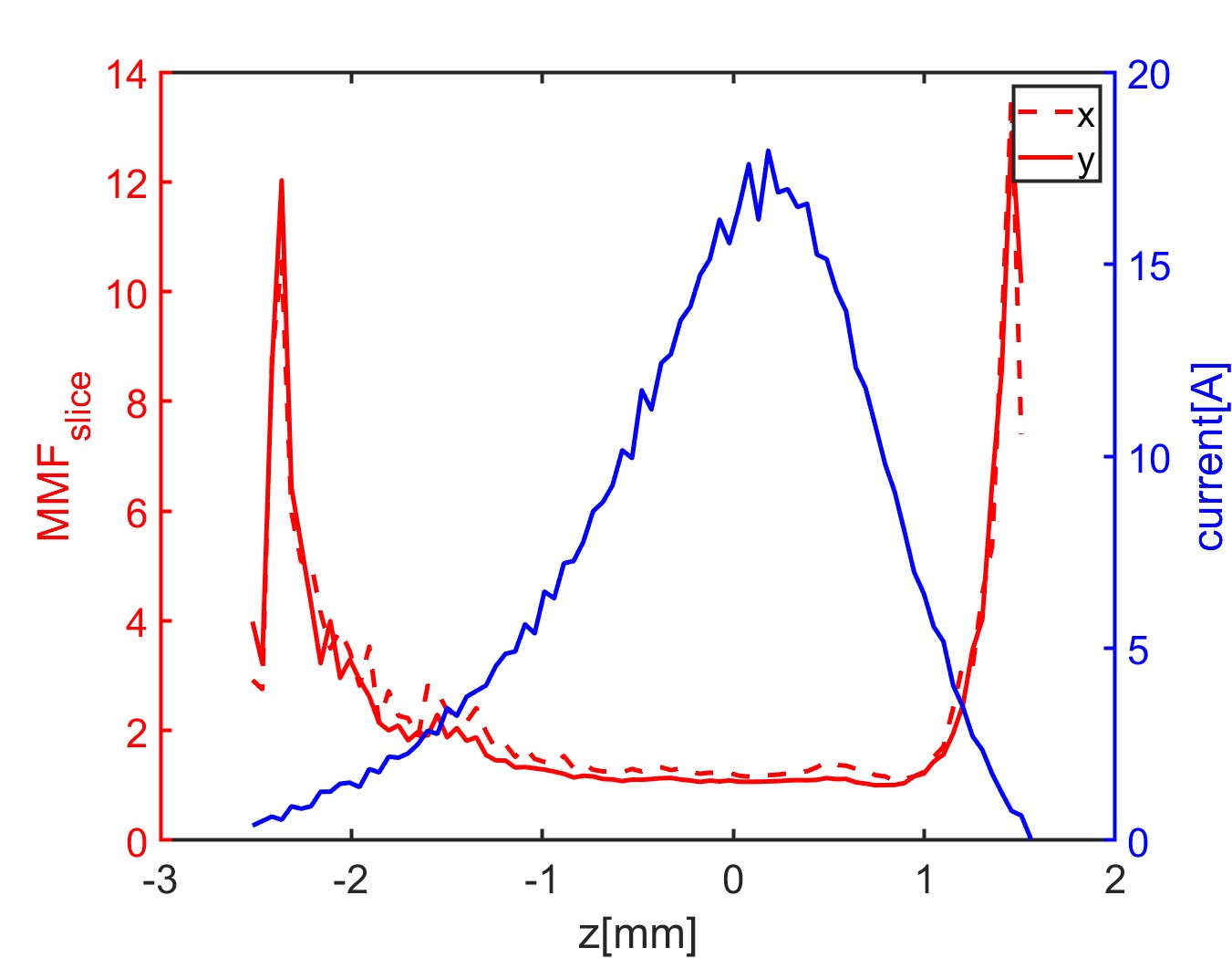}
    \includegraphics[width=.45\hsize]{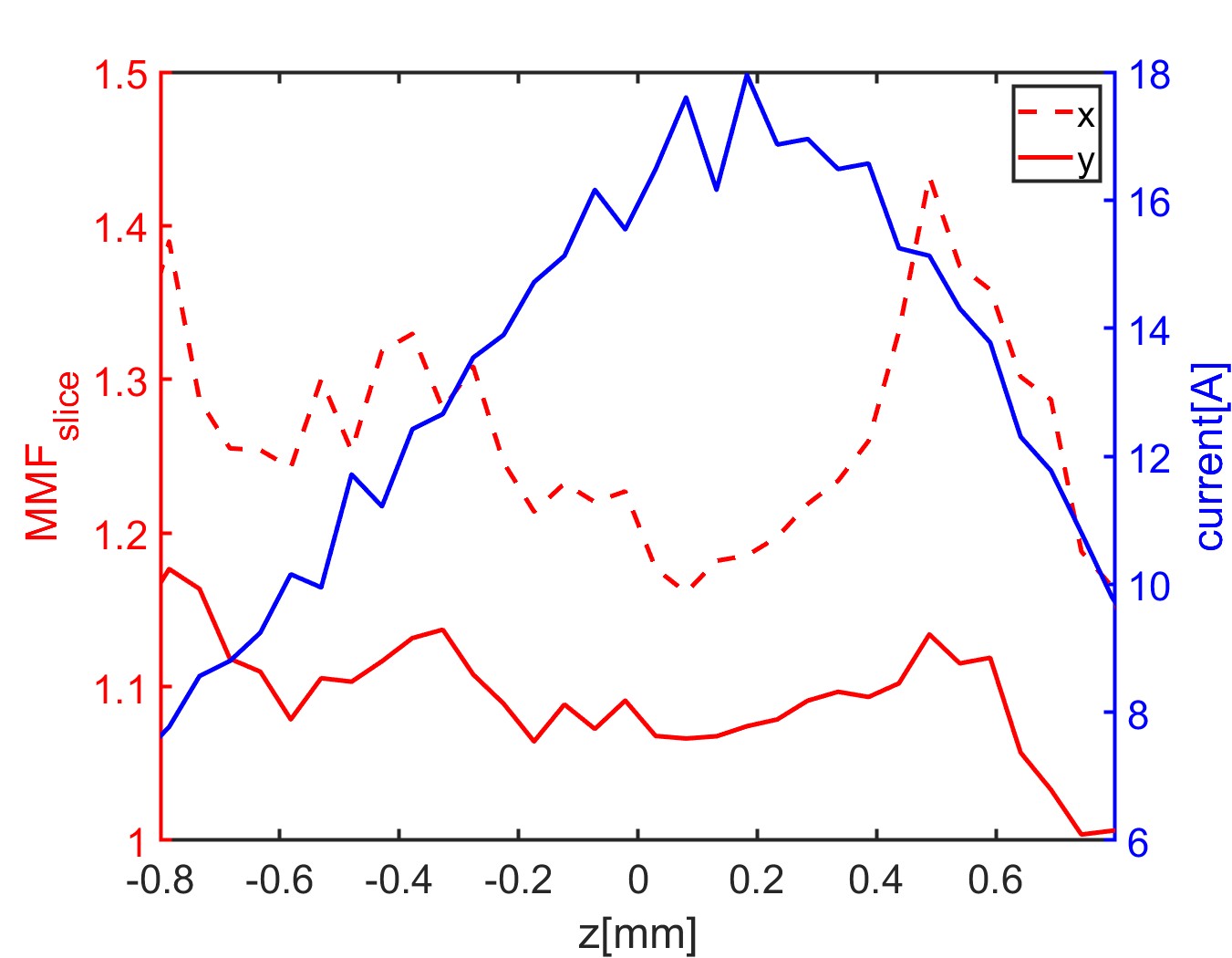}
    \\
    2(a) \hspace{3.2cm} 2(b)\hspace{4cm}\\
        \includegraphics[width=.45\hsize]{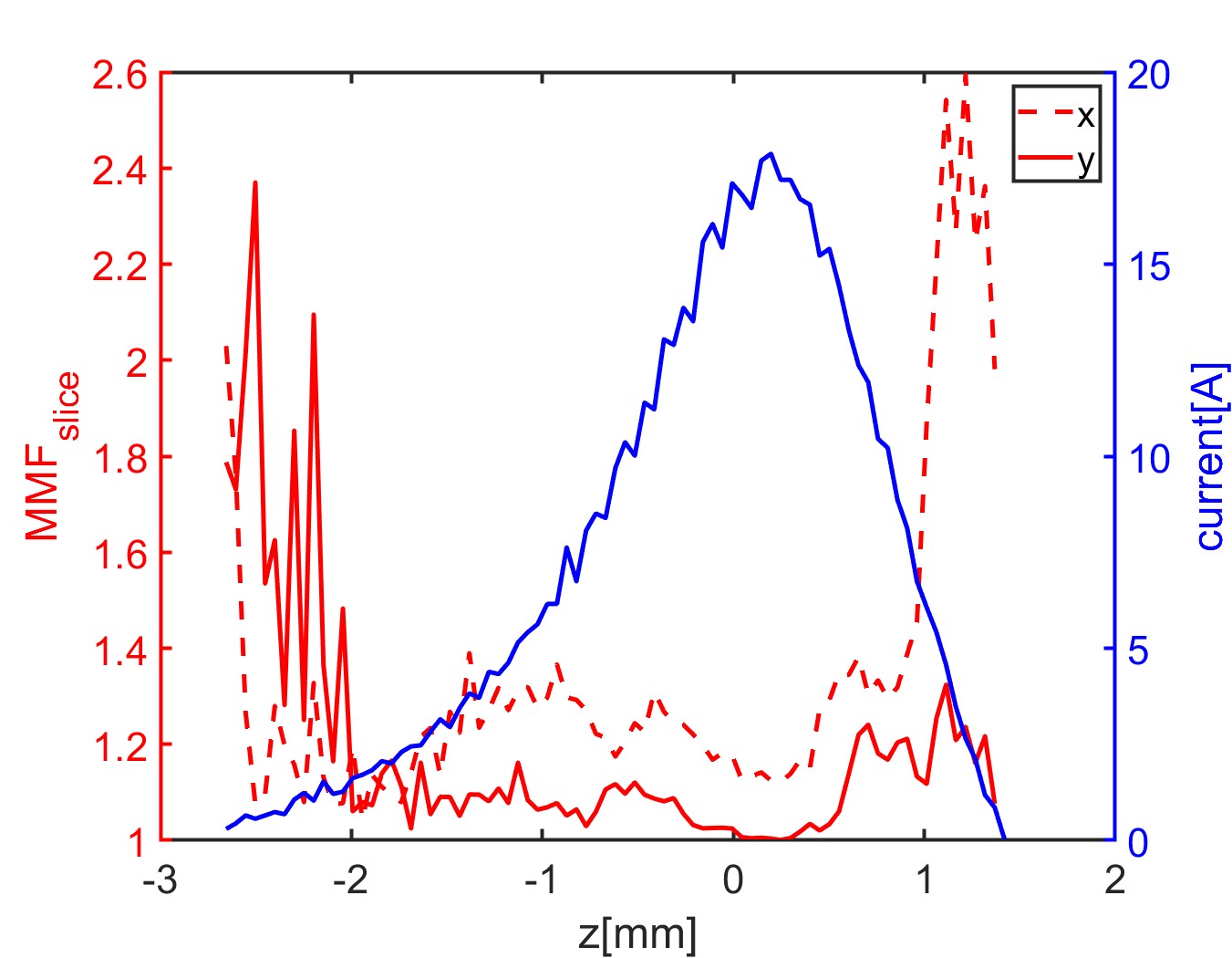}
    \includegraphics[width=.45\hsize]{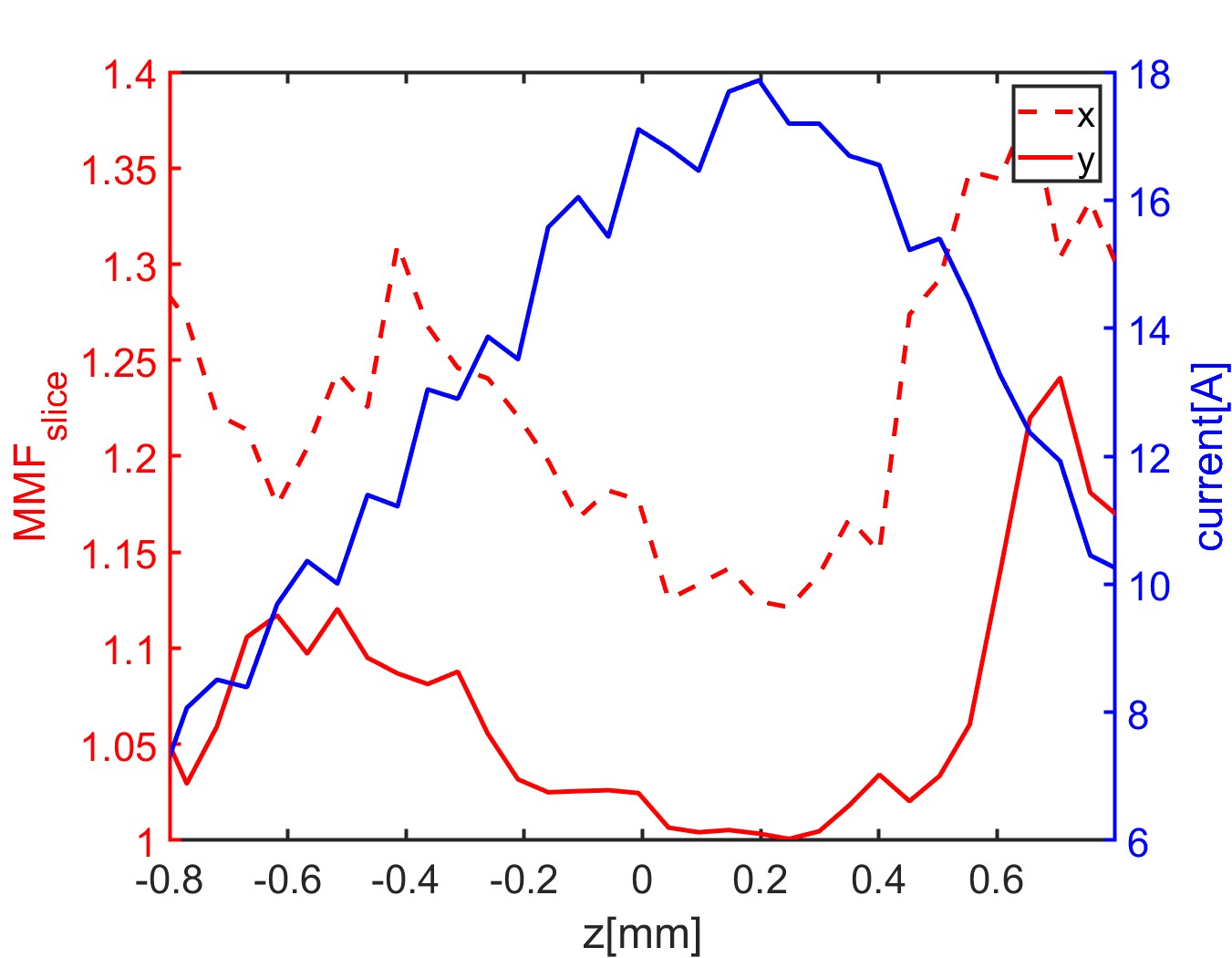}
    \\
    3(a) \hspace{3.2cm} 3(b)\hspace{4cm}\\
\caption{The slice mismath factor of the bunches. (1): the slice MMF at the exit of injector. (2): the slice MMF at the exit of match section. (3): the slice MMF at the exit of merger section. (a): the MMF of the complete bunches. (b): the MMF of the half-maximum intensity region of the current.  The blue line is the longitudinal profile of the bunches.}
    \label{fig:Phase_distribution}
\end{figure}
 For the non-ideal bunches, the mismatch occurs between the longitudinal bunch slices and the complete bunch. It can lead to a deviation that cannot be ignored in the Twiss functions across distinct slices. This deviation of the bunch slice from the ideal trajectory during propagation results in diminished beam quality\cite{PhysRevAccelBeams.20.013401}. Especially for slices with high current, a significant mismatch can lead to substantial degradation in beam quality. Furthermore, a matching section without collective effects is insufficient to mitigate the extent of mismatch, and once the beam attains high energy, it becomes increasingly challenging to counteract this effect. Related studies indicate that a match section, when operating within the space charge regime, has the potential to mitigate this mismatch\cite{miginsky2007minimization,miginsky2009emittance,miginsky2008space}. Then five quadrupole magnets were positioned in the matching section that connects the injector to the merger. The objective of this match section is to simultaneously manage the beam envelope within the merger section and mitigate mismatch. And the NSGA-II algorithm was employed to optimize the matching section\cite{inproceedings} To quantify the deviation, a well-known mismatch factor (MMF) was used to assess the mismatch between slices\cite{spence1991grid}:

 \begin{equation}
     \label{eqinj1}
 \begin{aligned}
 MMF&=\frac{1}{2}\mathrm{Tr\left(T^{-1}_{i}T_{0}\right)}\\
 \\
 \mathrm{T_i}&=\left(\begin{array}{ll}
 \beta_i &-\alpha_i\\
 -\alpha_i &\quad\gamma_i
 \end{array}\right)
 \end{aligned}
 \end{equation}
 $\mathrm{Tr(A)}$ the operation of taking the trace of a matrix A. $T_i$ refers to the Twiss matrix of the i-th slice of the beam. $T_0$ is the Twiss matrix of the complete bunch.  $\beta_i, \alpha_i$ and $\gamma_i$ are the Twiss functions of i-th slice. The mismatch disappears when the MMF equals 1. This factor is applicable to bunches with any distribution\cite{tang2008emittance} and can be easily demonstrated to be conserved under symplectic transformations. So, even though the bunch parameters from different optimization results may vary, this factor still effectively reflects the extent of mismatch within the bunch. During the optimization process, MMF values are weighted and summed, where the weighting factors are coefficients associated with the longitudinal distribution. Additionally, a greater weighting factor is assigned to MMF specifically in the horizontal plane. The results were shown in Fig.~(\ref{fig:Phase_distribution}). The MMF is reduced after passing through the match section and the merger section. Especially for the MMF of slices with large current, as illustrated in Fig.~(\ref{fig:Phase_distribution})(b). The MMF was significantly reduced at the exit of the match section, both in the horizontal and vertical planes. And it was further controlled after passing through the merger section. Especially in the vertical plane, the MMF for the slice adjacent to the slice with the maximum current was close to 1. Additionally, the MMF in the horizontal direction was reduced to approximately $30\%$ of its value at the exit of the injector.

\section{Discussion and Conclusion}

In this study, we propose a design for an ERL injector based on a VHF electron gun, demonstrating beam quality at the merger exit that is comparable to that of a conventional linear injector. This design fully satisfies the requirements for subsequent applications in superconducting linear accelerators and Free-Electron Lasers. However, current beam output records for VHF electron guns with high average beam currents remain limited. To address this issue, we are conducting tests to enhance the high average beam current output of VHF electron guns, aiming to increase the average beam current to the milliampere (mA) range.
Additionally, this article proposes utilizing three 3-cell superconducting cavities for injector acceleration. By integrating these three 3-cell cavities within the same module, we can meet the 10 MeV energy output requirement while conserving valuable injector space. For achieving high average beam currents in ERLs, semiconductor cathodes are considered the optimal choice. However, cesium telluride-based semiconductor cathodes rely on excitation by UV lasers, which face limitations due to power constraints associated with laser frequency doubling efficiency and reflectivity. Furthermore, the cathode plug tends to overheat from laser absorption, adversely affecting the quantum efficiency of the semiconductor cathode.
To overcome these challenges, we will employ a visible light excitation approach for the semiconductor cathode, complemented by a cathode plug made from highly thermally conductive metals such as molybdenum or copper.

A new method, termed $\zeta_{sc}$ functions, has been qualitatively proposed to predict the impacts of  LSC in non-zero dispersion regions, such as merger and Arc sections in space charge regime machines (see preliminary Arc design in \ref{A3}). Its simple formulation facilitates application in design work. The various merger types illustrated in Fig.~(\ref{fmer1}) are expressed as $\zeta_{sc}$ functions in \ref{A1}. For zigzag-type mergers, both $\zeta_{sc,1}$ and $\zeta'_{sc,1}$ approach zero, indicating superior effectiveness in suppressing first-order LSC effects compared to previous designs. The parameter $\delta_{{LSC}}$ necessitates the consideration of higher-order effects when the $R_{56}$ of the merger is non-zero. These effects can be estimated using $\zeta_{sc,i}$ and $\zeta'{sc,i}$ as defined in Eqs.~(\ref{eqmer6}). In constructing the objective function, only second-order effects (i.e., $i=2$) were incorporated. A coefficient proportional to $|R_{56}|$ was introduced before $\zeta_{sc,2}$. As $R_{56}$ approaches zero, this coefficient diminishes, leading $\delta_{{LSC}}$ to approach linearity and rendering the contribution of the $\zeta_{sc,2}$ term to the objective function negligible. This optimization function facilitated the design of a TBA merger with three quadrupole magnets. Comparative simulations demonstrated that the LSC mitigation capabilities of the TBA merger rival those of the zigzag merger, which was previously considered optimal but challenging to implement. Furthermore, the TBA merger exhibits enhanced tolerance for initial energy chirps due to its reduced $R_{56}$. Simulations reveal well-preserved emittance for initial chirps within the range of $[-10, 10]$, thereby enabling a broader optimization range for the acceleration phases of the injector cavity.
In future work, to predict the variance in emittance more accurately, the coefficients for the $i$-th $\zeta_{sc,i}$ will require further clarification. Additionally, under the paraxial approximation ($\sigma_x << \gamma \sigma_z$), a quantitative theory for estimating LSC effects in  beamlines with non-zero dispersion appears feasible. Research in this direction is currently ongoing, and the supplementation of related work is expected to advance the design efforts for merger sections and low-energy ERL Arc sections.
 \section{Appendix}
 \subsection{$\zeta_{sc}$ function for merger section as shown in Fig.~(\ref{fmer1})}

 \label{A1}
\begin{figure}
  \renewcommand{\thefigure}{A.1}
 
 \begin{center}
 \includegraphics[width=0.8\linewidth]{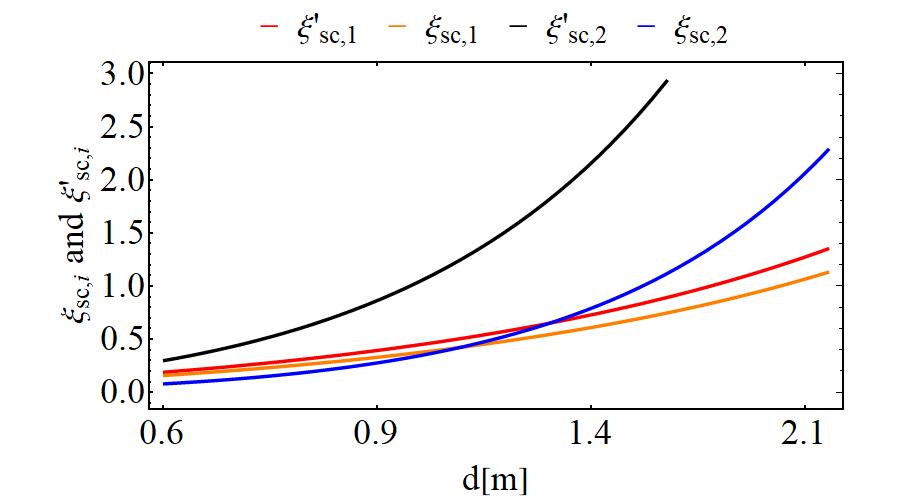}\\
 \includegraphics[width=0.8\linewidth]{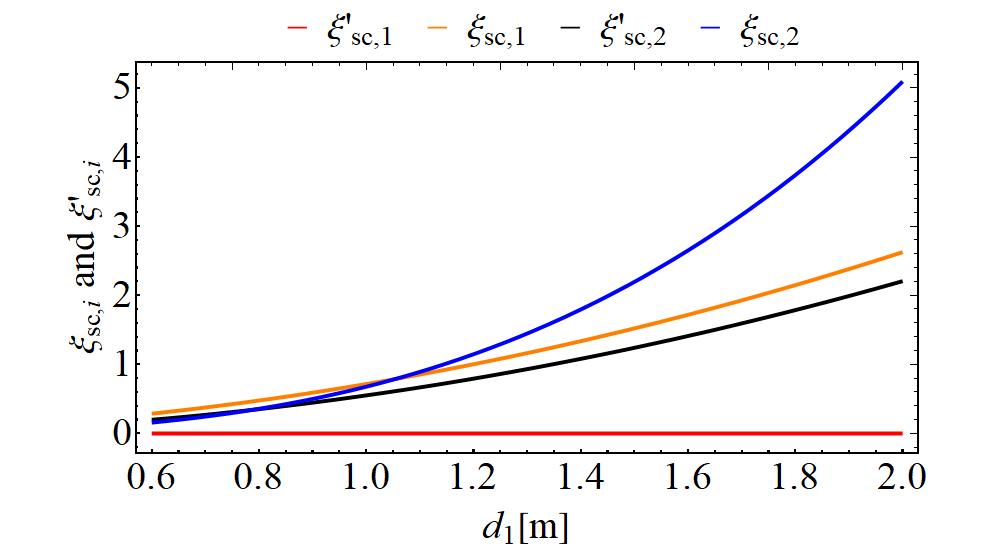}\\
 \includegraphics[width=0.8\linewidth]{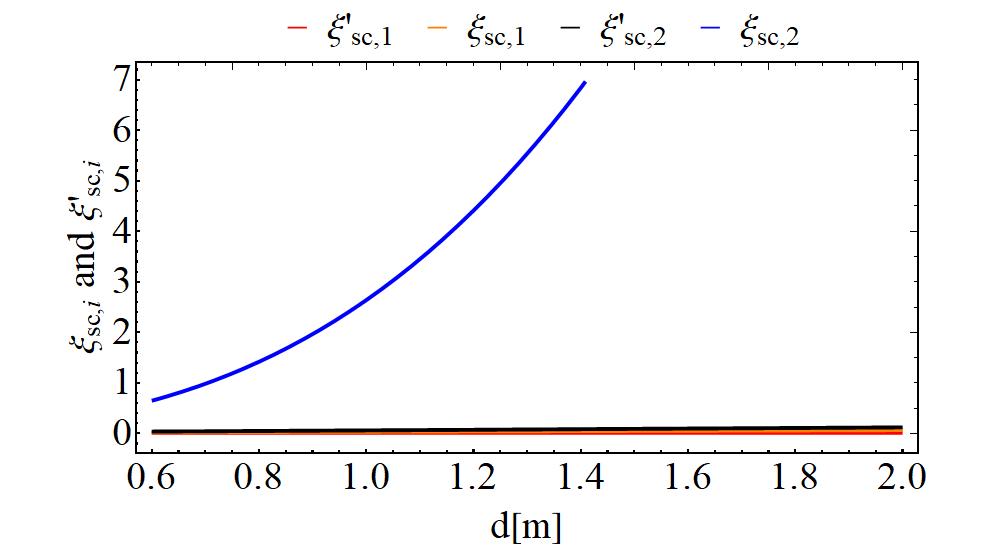}
 \caption{\label{famer}Top: The $\zeta_{sc}$ function for 3-B merger section. Middle: The $\zeta_{sc}$ function for zigzag merger section. Bottom: The $\zeta_{sc}$ function for zigzag merger section.}
\end{center}
 \end{figure}

 In this study, three types of merger sections were examined, namely the chicane merger, zigzag merger, and 3-B merger. To simplify the derivation process, all magnets were set to the same radius and maintaining the hard-edge approximation in this section. And the energy of the bunch in the merger is larger than 5$MeV$ so the impact of the relativistic factor on the transfer matrix was neglected. To make them achromatic, the following conditions must be met:
 \begin{equation}\tag{A.1}\label{a1}
 \begin{aligned}
 &d_2-2d_1=2R(1-\cos\theta)/\tan\theta\quad(zigzag)\\
 &d=R(\cot\theta_1-\cot\frac{\theta_2}{2}-2\csc\theta_1)\quad(3-B)
 \end{aligned}
 \end{equation}
 $\theta_1$ and $\theta_2$ are the angle of the first and the second bend for 3-B re-merger. $d_1$ and $d_2$ are the length of drifts after the first and the second bend in zigzag re-merger. All structures are symmetric about the midpoint. The angle of the first bend in chicane or 3-B re-merger is $16^{\circ}$ and $10^{\circ}$ in zigzag re-merger. The radius of bends is $0.5m$. The minimal acceptable length of the first drift is $0.6m$. 
 For the 3-B merger, we have:
 \begin{equation}\tag{A.2}\label{a2}
 \begin{aligned}
 R_{51,1}=-\sin\theta_1&,\quad R_{52,1}=-R(1-\cos\theta_1)\\
 R_{51,2}=&-3\sin\theta_1-2\sin\theta_2 \\ 
 R_{52,2}=R(-1+3\cos&\theta_1+2\cot\theta_1sin\theta_2-\csc\theta_1\sin\theta_2)
 \end{aligned}
 \end{equation}
 where $R_{5i,j}$ is the $R_{5i}$ after the $j$th bend in the re-merger. The $\zeta_{sc}$ function for 3-B merger is shown in top figure of Fig.~(\ref{famer})

 In a 3-B merger, the value of the $\zeta_{sc}$ function increases as the mid-bend angle decreases. Therefore, compared to a 3-B merger with three identical bends, increasing the angle of the mid-bend can reduce the impact of LSC on the transverse plane. With specific magnetic field strengths in the bends, a 3-B merger with quadrupole magnets can adjust the $R_{56}$ and $\zeta_{sc}$ functions. After experimenting with several designs,  it was observed that compared to the best case scenario described in this section (i.e. $\theta_2=0.4$), the $\zeta_{sc}$ functions did not significantly decrease. However, incorporating additional quadrupole magnets can enable the merger to achieve specific $R_{56}$ with lower $\zeta_{sc}$ functions, which is crucial when dealing with bunches that have varying energy chirps. It's important to note that when the angle is large as the length of the drift decreases, the LSC within the bend cannot be neglected. This is due to the fact that $\int_{B} R_{5i} d s <<\int_{all} R_{5i} d s$ no longer holds in these cases.


 In this study, C-type chicane with four identical bends was considered. For the chicane merger:
 \begin{equation}\tag{A.3}\label{a3}
 \begin{aligned}
 R_{51,1}=-\sin\theta&,\quad R_{52,1}=-R(1-\cos\theta)\\
 R_{51,2}=&0 \\ 
 R_{52,2}=2 \sec \theta^2 \sin\frac{\theta}{2}&(d_1 \cos \frac{\theta}{2}+R(- \sin \frac{\theta}{2}+ \sin \frac{3 \theta}{2}))\\
 R_{51,3}=&\sin\theta\\
 R_{52,3}=\frac{1}{2} \sec \theta^2 \sin\frac{\theta}{2}&(2 \left(4d_1+d_2\right)\cos \frac{\theta}{2}+d_2\cos \frac{3\theta}{2}\\
 +d2\cos \frac{5\theta}{2}-&2R\sin \frac{\theta}{2}+5R\sin \frac{3\theta}{2}+3R\sin \frac{5\theta}{2})\\
 \end{aligned}
 \end{equation}

 Different from other types of mergers discussed in this section, there are two independent variables for chicane merger: the first drift $d_1$ and the second drift $d_2$. Based on the analytical expression of the $\zeta_{sc}$ function, it decreases with a reduction in $d_2$ for any given $d_1$. The variation of the $\zeta_{sc}$ function with changes in $d_1$ is shown in the middle figure of Fig.~(\ref{famer}) when the length of $d_2$ is zero. The $\zeta'_{sc,1}$ is equal to zero for any length of $d_1$. C-type chicane with asymmetric structure was also considered in our study, but it leaded to a better case.
 

 For the zigzag merger:
 \begin{equation}\tag{A.4}\label{a4}
 \begin{aligned}
 R_{51,1}=-\sin\theta&,\quad R_{52,1}=-R(1-\cos\theta)\\
 R_{51,2}=&\sin\theta \\ 
 R_{52,2}=\frac{1}{2} \sec \theta^2 \sin\frac{\theta}{2}&\left(8 d_1 \cos \frac{\theta}{2}+R\left(-2 \sin \frac{\theta}{2}+5 \sin \frac{3 \theta}{2}+3 \sin \frac{5\theta}{2}\right)\right)\\
 R_{51,3}=&-\sin\theta\\
 R_{52,3}=\frac{1}{2} \sec \theta^2 \sin\frac{\theta}{2}&\left(16 d_1 \cos \frac{\theta}{2}+R\left(-11 \sin \frac{\theta}{2}+5 \sin \frac{3 \theta}{2}+3 \sin \frac{5\theta}{2}\right)\right)\\
 \end{aligned}
 \end{equation}

 The $\zeta_{sc}$ function for zigzag merger is shown in the bottom figure of Fig.~(\ref{famer}). $\zeta_{sc,1}$, $\zeta'_{sc,1}$ and $\zeta'_{sc,2}$ close to zero for any length of $d_1$ in all region. For the cases that the variance in bunch length $\Delta\sigma_{z}$ in merger is much smaller than initial bunch length, so-called 'frozen cases' in ref\cite{Venturini2015}, the $k_2(z)$ is much smaller than $k_1(z)$ in Eqs.~(\ref{eqmera2}). It is leaded to the linear manner for the variance in energy spread cased by LSC.  The $\zeta_{sc,2}$ can be neglected since the weight of  $\zeta_{sc,1}$ and $\zeta'_{sc,1}$ are significantly greater than that of $\zeta_{sc, 2}$ and $\zeta'_{sc,2}$ in these cases. 

 In the case of the DBA or a two-bend dog-leg merger as outlined in Ref\cite{RossettiConti2023}, the $\zeta_{sc}$ functions are determined solely by the characteristics of the first bend in the re-merger section and by the overall length of the merger section. The $R_{5i}$ and $\zeta_{sc}$ functions can be expressed as

 \begin{equation}\tag{A.5}\label{a4}
 \begin{aligned}
 R_{51,1}=-\sin\theta&,\quad R_{52,1}=-R(1-\cos\theta)\\
 \zeta_{sc,1}=R_{52}\left(l_{all}-2l_B\right)&, \quad
 \zeta'_{sc,1}=R_{51}\left(l_{all}-2l_B\right)\\
 \zeta_{sc,2}=R_{52}&\left(l_{all}-l_B\right)^2-l_b^2\\
 \zeta'_{sc,2}=R_{51}&\left(l_{all}-l_B\right)^2-l_b^2\\
 \end{aligned}
 \end{equation}
 where $l_b$ is the length of bends in DBA merger.

\subsection{A preliminary design of Arc sections with low energy.}

\label{A3}

When the machine operates at lower energies, the effects of LSC cannot be neglected. Arc sections characterized by a large value of $\zeta_{sc}$ functions would result in a dilution of the transverse emittance. A similar method applied to the merger design is utilized for the Arc sections. And the variance in bunch length is considered to be close to zero in this preliminary design. Consequently, only the first order of $\zeta_{sc}$ functions was taken into account in the design work. The Arc sections adopted a symmetrical TBA structure, which represents the simplest configuration capable of minimizing the $\zeta_{sc}$ functions close to zero in the first order. The total deflection angle of the TBA structure amounts to $180^{\circ}$. Another unavoidable issue is that there are five constraints for this structure to ensure the variance in bunch length is sufficiently small: 1) achromatic, 2) symplectic condition for transfer matrix, 3)$\&$4) $\zeta_{sc}=0$, 5)$R_{56}$ close to zero. And there are four degrees of freedom after the total deflection angle and radius of the bending magnets are determined, which means some of the above five conditions cannot be satisfied.  

\begin{figure} 
  \renewcommand{\thefigure}{C.1}
 
 \begin{center}
 \includegraphics[width=0.8\linewidth]{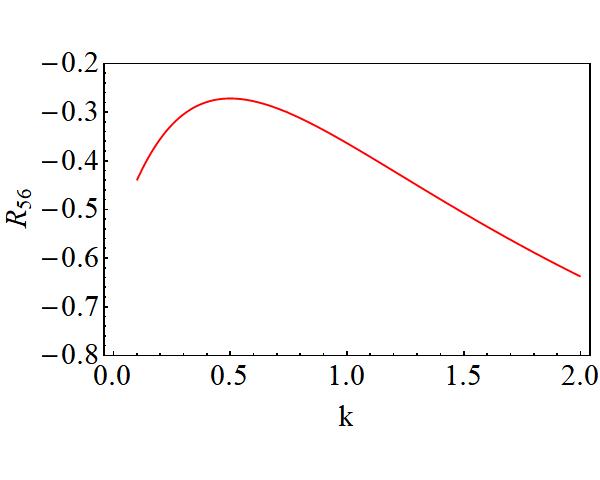}
 \caption{\label{r56vsk}The $R_{56}$ varies with the angle ratio $k$. The angle ratio between the middle bend and the side bend is $2k:1$.}
\end{center}
 \end{figure}

In this context, we aim to concentrate specifically on the effectiveness of the $\zeta_{sc,1}$ functions, hence conditions 1) through 4) are prioritized. Although it is impossible to make $R_{56}$ approach zero in the design, we can still minimize the variation in bunch length by choosing an appropriate angle ratio and setting the initial  energy chirp of bunch to zero. The $R_{56}$ varies with the angle ratio $k$ between the half of the middle bend and the side bend as shown in Fig.\~(\ref{r56vsk}) and $R_{56}$ is closer to zero in the case where $k$ equals to 0.5. The bunch parameters are referenced in Table~(\ref{tabmer2}), except that the energy was set to 35 MeV. The simulation results were shown in Table~(\ref{tab_arc_result}). The growth in emittance remains within 0.1 $\mathrm{\mu mrad}$ even after considering the space charge effects. The growth in emittance caused by space charge was limited to within $5\%$. At the same time, the impact of CSR on the transverse distribution of the bunch was also effectively constrained. This is because under the steady-state approximation, the displacements in the $x$-plane caused by CSR can be expressed as
\begin{equation}\tag{C.1}\label{c1}
\begin{aligned}
& \hat{x}_{_{CSR}}(z)=\int_{B} \delta_{_{CSR}}^{\prime}(z) \cdot R_{52} d s \\
& \hat{x}_{_{CSR}}^{\prime}(z)=\int_{B} \delta_{_{CSR}}^{\prime}(z) \cdot R_{51} d s .
\end{aligned}
 \end{equation}

\begin{table}[!htb]
    \centering
\caption{The simulation results of Arc sections. 'no SC' the simulatio without collective effects. 'SC$\&$CSR' the simulations with space charge and steady-state CSR.}
\label{tab_arc_result}
\begin{tabular}{cccc}
\hline\hline\\ &\quad\quad no SC &\quad\quad SC &\quad\quad SC\&CSR \\
\\ \hline\\$\Delta \varepsilon_x / \varepsilon_0$ &\quad\quad $0 \%$ &\quad\quad $0.4 \%$ &\quad\quad $2 \%$ \\ \\
\hline\hline
\end{tabular}
\end{table}
It has a similar form as Eqs.~(\ref{eqmer3}), with the distinction that these integral are applied within the confines of the Bends. Although it is not possible to make the right-hand side of Eqs.~(\ref{c1}) equal to zero, setting the $R_{5i}$ values of the middle bend's sides to be opposites can cancel out the transverse displacements caused by CSR in different bends of the TBA. In practical designs, the transverse emittance is better preserved by comparing the impacts of the two types of collective effects on the bunch. Furthermore, to make $R_{56}$ of the TBA Arc section approach zero, two feasible strategies are the use of an asymmetric TBA or a more complex Multi-Bend Achromat(MBA) structure. Additional degrees of freedom could allow all five previously mentioned conditions to be satisfied. Related work will be supplemented in the future.

\section*{CRediT authorship contribution statement}
\textbf{Xiuji Chen:} Methodology, Investigation, Software, Data and analysis curation, Project administration, Writing - original draft, Writing - review \& editing. \textbf{Zipeng Liu:}Methodology, Investigation, Software, Data and analysis curation, Project administration, Writing - original draft, Writing - review \& editing. \textbf{Xuan Huang:} Data. \textbf{Si Chen:} Review. \textbf{Duan Gu:} Review.\textbf{Houjun Qian:} Review. \textbf{Haixiao Deng:} Funding acquisition, Review. \textbf{Dong Wang:} Review.

\section*{Declaration of competing interest}
The authors declare that they have no known competing financial interests or personal relationships that could have appeared to influence the work reported in this paper.

\section*{Acknowledgement}
This work was supported by the CAS project for Young Scientists in Basic Research (YSBR-042), The National Natural Science Foundation of China (12125508, 11935020), Program of Shanghai Academic Technology Research Leader (21XD1404100), and Shanghai Pilot Program for Basic Research Chinese Academy of Sciences, Shanghai Branch (JCYJ-SHFY-2021-010)


\bibliography{mybibfile}

\end{document}